\documentclass[12pt]{iopart}
\usepackage{iopams}
\usepackage{graphicx}
\expandafter\let\csname equation*\endcsname\relax
\expandafter\let\csname endequation*\endcsname\relax
\usepackage{amsmath}
\usepackage{amssymb}
\usepackage{cite}

\begin{document}

\title[Cover times of random walks on random regular graphs]
{Analytical results for the distribution of cover times 
of random walks on random regular graphs
}

\author{Ido Tishby, Ofer Biham and Eytan Katzav}
\address{Racah Institute of Physics, 
The Hebrew University, Jerusalem 9190401, Israel.}
\eads{\mailto{ido.tishby@mail.huji.ac.il}, \mailto{biham@phys.huji.ac.il}, 
\mailto{eytan.katzav@mail.huji.ac.il}}

\begin{abstract}

We present analytical results for the distribution of
cover times of random walks (RWs) on random regular graphs
consisting of $N$ nodes of degree $c$ ($c \ge 3$).  
Starting from a random initial node at time $t=1$, 
at each time step $t \ge 2$ an RW hops 
into a random neighbor of its previous node.
In some of the time steps the RW may visit a new, yet-unvisited node,
while in other time steps it may revisit a node that has already been visited before.
The cover time $T_{\rm C}$ is the number of time steps required for the RW to visit
every single node in the network at least once.
We derive a master equation for the distribution
$P_t(S=s)$ of the number of distinct nodes $s$ visited by an RW up to time $t$
and solve it analytically.
Inserting $s=N$ we obtain the cumulative distribution of cover times, namely
the probability 
$P(T_{\rm C} \le t) = P_t(S=N)$
that up to time $t$ an RW will visit all the
$N$ nodes in the network.
Taking the large network limit, we show that
$P(T_{\rm C} \le t)$
converges to a Gumbel distribution.
We calculate the distribution of partial cover (PC) times
$P( T_{{\rm PC},k} = t )$,
which is the probability that at time $t$ an RW will complete visiting
$k$ distinct nodes.
We also calculate the distribution of random cover (RC) times
$P( T_{{\rm RC},k} = t )$,
which is the probability that at time $t$
an RW will complete visiting all the nodes in a subgraph of $k$ randomly pre-selected nodes at least once.
The analytical results for the distributions of cover times are 
found to be in very good agreement with the results 
obtained from computer simulations.

\end{abstract}

%\pacs{05.40.Fb, 64.60.aq, 89.75.Da}


%\submitto{\jpa (\today)}
%\date
\maketitle

\section{Introduction}

Random walk (RW) models 
\cite{Spitzer1964,Weiss1994}
were studied extensively 
in different geometries, including
continuous space
\cite{Lawler2010b}, 
regular lattices 
\cite{Lawler2010a},
fractals 
\cite{ben-Avraham2000}
and 
random networks
\cite{Noh2004,Burioni2005,Masuda2017,Dorogovtsev2010,Estrada2011}.
These models are useful for the analysis of a large variety of
stochastic processes 
such as diffusion
\cite{Berg1993,Ibe2013},
polymer structure 
\cite{Edwards1965,Fisher1966,Degennes1979} 
and random search
\cite{Evans2011,Lopez2012}.
In the context of complex networks
\cite{Havlin2010,Newman2010,Estrada2015}  
they provide useful insight on the
spreading of rumours, opinions and infections
\cite{Pastor-Satorras2015,Barrat2012}.
Consider an RW on a random network 
that starts at time $t=1$ from a random initial node $x_1$. 
At each time step $t \ge 2$ the RW hops
randomly to one of the neighbors of its previous node.
The RW thus generates a trajectory of the form
$x_1 \rightarrow x_2 \rightarrow \dots \rightarrow x_t \rightarrow \dots$,
where $x_{t}$ is the node visited at time $t$.
In some of the time steps the RW hops into nodes that
have not been visited before, while
in other time steps it hops into nodes that have
already been visited at an earlier time.
Since RWs on random networks may visit some of the nodes more than once,
the number of distinct nodes visited up to time $t$  
is typically smaller than $t$.
The mean number $\langle S \rangle_t$ of distinct nodes visited by an RW 
on a random network up to time $t$ was recently studied 
\cite{Gallos2004,Debacco2015,Tishby2021b}.
It was found that 
in the infinite network limit it scales linearly with $t$, namely
$\langle S \rangle_t \simeq r t$, 
where the coefficient
$r<1$ depends on the network topology.
These scaling properties resemble those obtained
for RWs on high dimensional lattices
\cite{Dvoretzky1951,Vineyard1963,Montroll1965,Finch2003}, 
Bethe lattices and Cayley trees
\cite{Mezard1987,Cassi1989,Rogers2008,Martin2010}.
For finite networks of size $N$, the linear relation between $\langle S \rangle_t$ and $t$
holds as long as $t \ll N$. 
At longer times, 
the probability of an RW to enter a yet-unvisited node gradually decreases.
As a result, $\langle S \rangle_t$ eventually saturates,
converging towards $N$ at $t \rightarrow \infty$. 

RWs on random networks exhibit a variety of first passage processes
\cite{Redner2001}, which take place over a broad range of time scales. 
These include the first hitting time $T_{\rm FH}$,
which is the first time at which an RW steps into a
node that has already been visited before
\cite{Herrero2003,Herrero2005b,Tishby2016,Tishby2017,Tishby2017b,Tishby2017c,Tishby2021a}.
The mean first hitting time satisfies
$\langle T_{\rm FH} \rangle  \sim  \min \{ c, \sqrt{N} \}$,
where $N$ is the network size and $c$ is the mean degree
\cite{Tishby2017,Tishby2021a}.
This implies that $\langle T_{\rm FH} \rangle \sim c$ in dilute networks 
and $\langle T_{\rm FH} \rangle  \sim  \sqrt{N}$ in dense networks.
The first passage time $T_{\rm FP}$
is the first time at which an RW 
starting from a given initial node $i$
visits a given target node $j$
\cite{Sood2005}.
In a finite network that consists of a single connected component,
the mean first passage time satisfies
$\langle T_{\rm FP} \rangle \sim N$.
A special case of the first passage time is the first return time $T_{\rm FR}$,
which is the first time at which an RW returns to the initial node $i$
\cite{Masuda2004,Tishby2021b}.
The mean first return time satisfies
$\langle T_{\rm FR} \rangle \sim N$
\cite{Dorogovtsev2010,Tishby2021b}.
Interestingly, this result can be obtained directly from the Kac lemma,
which empolys general properties of discrete stochastic processes
\cite{Kac1947}.

The cover time is the number of time steps required
for an RW to visit every single node in a finite network
of size $N$ (consisting of a single connected component) at least once
\cite{Aleliunas1979,Aldous1989,Aldous1989b,Broder1989,Kahn1989,Lovasz1993,Feige1995a,Feige1995b,Jonasson1998,Abdullah2012,Abdullah2012b,Cooper2007,Cooper2007b,Cooper2008,Zlatanov2009,Cooper2014,Maier2017,Frieze2020}.
The cover time is relevant to a broad range of random search processes.
These include search processes involving multiple targets in which all the targets
need to be found. More specifically, in case that the number of targets is unknown
one needs to perform an exhaustive search in which all the nodes in the network 
must be visited.
Examples of such situations include the chase of pathogens by immune-system cells
\cite{Heuze2013},
foragers searching for food
\cite{Benichou2014,Chupeau2016a,Chupeau2016b}
and cleaning and demining processes in random environments.
In the special case of $c=N-1$ (complete graph)
the cover-time problem 
is analogous to the coupon collector problem
\cite{Lovasz1993,Holst2001,Neal2008}. 
Another interesting relation is to the random deposition model
\cite{Barabasi1995},
in the sense that the cover time corresponds to the
time at which the last
exposed substrate site is covered by a particle.

The mean cover time 
$\langle T_{\rm C} \rangle$
of RWs on general graphs has been studied extensively
since the late 1980s.
In particular, upper and lower bounds for 
$\langle T_{\rm C} \rangle$
for specific families of graphs were derived.
Regarding the lower bound, it was shown that for any
connected graph of $N$ nodes, the mean cover time satisfies
$\langle T_{\rm C} \rangle > [1 + {o}(1)] N \ln N$
\cite{Feige1995a}.
Actually, this lower bound was proven more directly,
and is believed to be tighter,
in the context of Erd{\H o}s-R\'enyi (ER) networks
\cite{Jonasson1998}.
In fact, in sparse ER networks, where $c < \ln N$
the mean cover time scales like 
$\langle T_{\rm C} \rangle \sim N (\ln N)^2$,
while in dense ER networks, where $c>\ln N$
the mean cover time scales like
$\langle T_{\rm C} \rangle \sim N  \ln N$
\cite{Cooper2008,Barlow2011}.
As for the upper bound, 
it was shown that for any connected graph
of $N$ nodes,
$\langle T_{\rm C} \rangle \le 4 N^3 / 27 + \mathcal{O}(N^{5/2})$
\cite{Feige1995b}.
For any regular graph
(a graph in which all the nodes are of the same degree)
that consists of a single connected component 
it was shown that
$\langle T_{\rm C} \rangle \le 8 N^2$
\cite{Kahn1989}.
Finally, for random regular graphs (RRGs) consisting of $N$ nodes of degree $c$ it was shown that
in the asymptotic limit
\cite{Cooper2005}

\begin{equation}
\langle T_{\rm C} \rangle \simeq \frac{c-1}{c-2} N \ln N.
\label{eq:TC1}
\end{equation}

\noindent
%Note that this result echoes the lower bound mentioned above for ER
%networks
%\cite{Jonasson1998},
%namely that $\langle T_{\rm C} \rangle$ is larger than $N \ln N$ by a 
%multiplicative factor which is slightly larger than $1$.
However, very little is known about the distribution of cover times $P(T_{\rm C}=t)$.
%We are aware of a single previous attempt to provide an analytical method to calculate
For graphs of small size the distribution of cover times 
can be calculated using the method of Ref.
\cite{Zlatanov2009}.
This method yields an approximation scheme that can be used for larger graphs,
and whose computational complexity scales like $\mathcal{O}(2N)$.
Nevertheless, no explicit analytical results for the distribution of cover times
are available for random graphs.

In this paper we present analytical results for 
the distribution of cover times
of RWs on RRGs
consisting of $N$ nodes
of degree $c \ge 3$.
To this end, we 
derive a master equation for the distribution
$P_t(S=s)$ of the number of distinct nodes $s$ visited by an RW up to time $t$. 
Using a generating function formalism, we 
solve the master equation and
obtain a closed-form analytical expression
for $P_t(S=s)$. 
Applying this result to the special case of $s=N$, we obtain
the cumulative distribution of cover (C) times,
which is given by 
$P( T_{\rm C} \le t)=P_t(S=N)$. 
%Using a generating function formulation, 
We also calculate the mean and variance of the distribution of
cover times.
Taking the large network limit, we show that
$P(T_{\rm C} \le t)$
follows a Gumbel distribution.
We also study two interesting generalizations of the cover time:
the partial cover (PC) time 
$T_{{\rm PC},k}$, which is the
time it takes an RW to visit $k$ distinct nodes and the random cover (RC) time 
$T_{{\rm RC},k}$,
which is the time it takes an RW to cover a set of $k$ randomly pre-selected nodes
\cite{Chupeau2015,Nascimento2001,Coutinho1994}.
The analytical results for the distributions of cover times are 
found to be in very good agreement with the results 
obtained from computer simulations.

The paper is organized as follows.
In Sec. 2 we briefly describe the RRG.
In Sec. 3 we present the random walk model.
In Sec. 4 we derive the master equation for $P_t(S=s)$.
In Sec. 5 we present the solution of the master equation in
the infinite network limit.
In Sec. 6 we present the solution of the master equation 
for finite networks.
In Sec. 7 we calculate the mean and variance of $P_t(S=s)$.
In Sec. 8 we calculate the distribution of cover times.
In Sec. 9 we calculate the mean cover time.
In Sec. 10 we calculate the variance of the distribution of cover times.
In Sec. 11 we present the distribution of partial cover times and in
Sec. 12 we consider the distribution of random cover times.
The relation between these two distributions is discussed in Sec. 13.
The results are discussed in Sec. 14 and summarized in Sec. 15.
In Appendix A we 
present the solution of the master equation for  $P_t(S=s)$.
In Appendix B we consider the distribution
$P(T=t|s)$ that an RW has pursued $t$ times steps
given that it has visited $s$ distinct nodes.
In Appendix C we calculate the moments of $P_t(S=s)$.
In Appendix D we calculate the generating function of the
distribution of cover times.

\section{The random regular graph}

A random network (or graph) consists of a set of $N$ nodes that
are connected by edges in a way that is determined by some
random process.
For example, in a configuration model network the degree of each node is 
drawn independently from a given degree distribution $P(k)$ and
the connections are random and uncorrelated
\cite{Molloy1995,Molloy1998,Newman2001}.
Configuration model networks belong to the class of small-world networks
in which the mean distance $\langle L \rangle$ between pairs of random nodes scales 
logarithmically with the network size, namely $\langle L \rangle \sim \ln N$
\cite{Newman2001}.
The RRG is a special case of a configuration 
model network, in which the degree distribution is a degenerate
distribution of the form 
$P(k)=\delta_{k,c}$, namely
all the nodes are of the same degree $c$.
Here we focus on the case of $c \ge 3$,
in which for a sufficiently large value of $N$ the RRG consists of a single connected component
\cite{Bollobas2001}.
In the infinite network limit the RRG exhibits a tree
structure with no cycles. 
Thus, in this limit it coincides with a Bethe lattice whose coordination number is equal to $c$
\cite{Bethe1935}.
In contrast, RRGs of a finite size exhibit a local tree-like structure,
while at larger scales there is a broad spectrum of cycle lengths.
In that sense RRGs differ from Cayley trees
\cite{Cayley1878}, which maintain their
tree structure by reducing the most peripheral nodes to leaf nodes of degree $1$.

A special property of RRGs is that there is a great deal of uniformity in the local
neighborhood of all nodes in the network. This property makes it an ideal model
for mean-field analysis, which often provides exact results.
For example, the distribution of shortest path lengths (DSPL) of RRGs  
\cite{Hofstad2005,Nitzan2016} 
as well as the distribution of shortest cycles
\cite{Bonneau2017}
are known exactly.

A convenient way to construct an RRG
of size $N$ and degree $c$
is to prepare the $N$ nodes such that each node is 
connected to $c$ half edges or stubs
\cite{Newman2010,Dorogovtsev2003}.
At each step of the construction, one connects a random pair of stubs that 
belong to two different nodes $i$ and $j$ 
that are not already connected,
forming an edge between them.
This procedure is repeated until all the stubs are exhausted.
The process may get stuck before completion in case that
all the remaining stubs belong to the
same node or to pairs of nodes that are already connected.
In such case one needs to perform some random reconnections
in order to complete the construction.

\section{The random walk model}

Consider an RW on an RRG of degree $c \ge 3$ and size $N$.
At each time step the RW hops from its current node to one of its neighbors,
such that the probability of hopping to each neighbor is $1/c$.
For sufficiently large $N$ the RRG consists of a single connected
component, thus an RW starting from any initial node
can reach any other node in the network.
In the long time limit $t \gg N$ an RW on an RRG visits all the nodes with the same frequency,
namely on average each node is visited once every $N$ steps.
However, over shorter periods of time there may be large fluctuations such that
some nodes may be visited several times
in a given time interval
while other nodes are not visited at all.

\begin{figure}
\centerline{
\includegraphics[width=5.0cm]{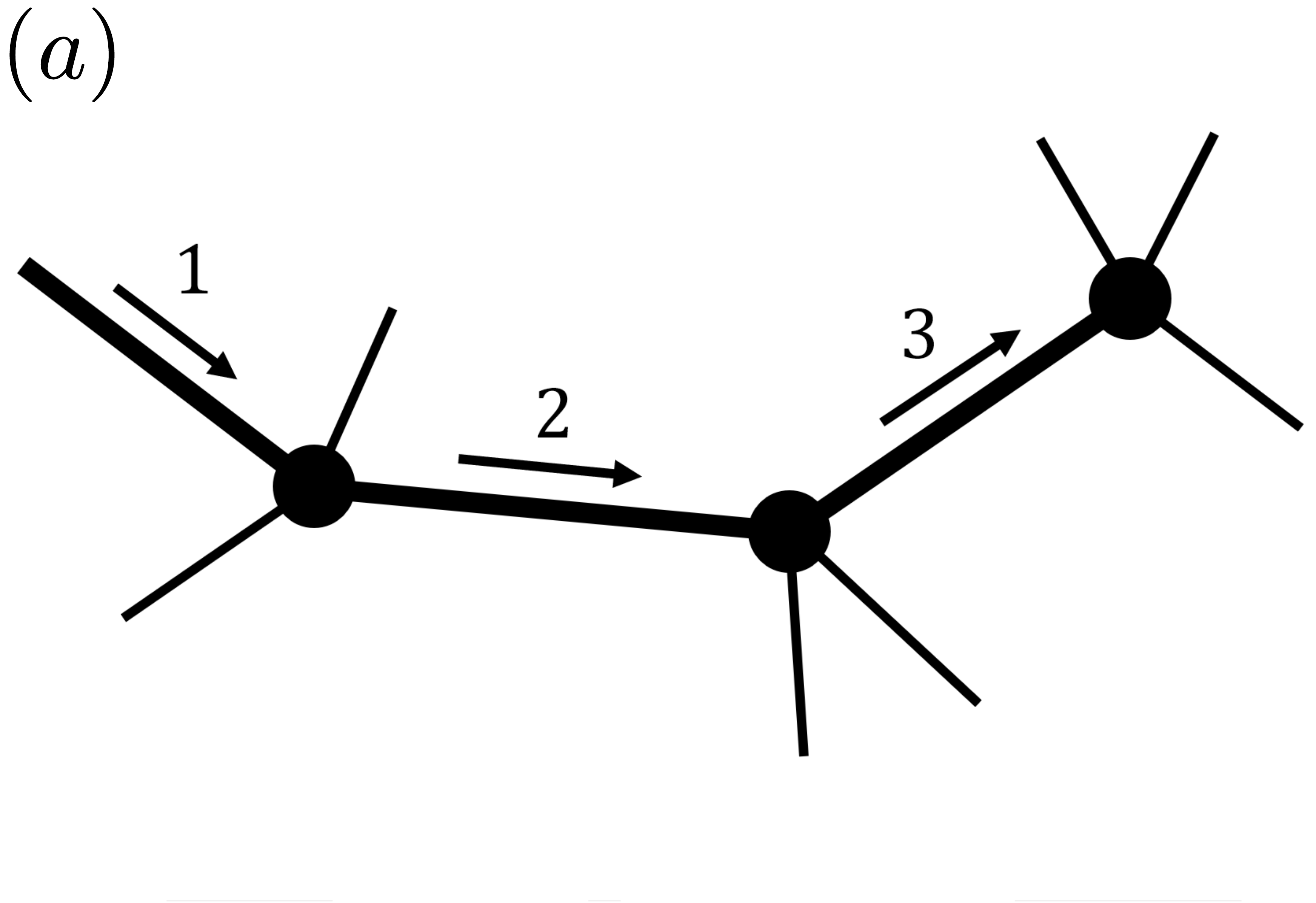}
\hspace{2cm}
%}
%\centerline{
\includegraphics[width=5.0cm]{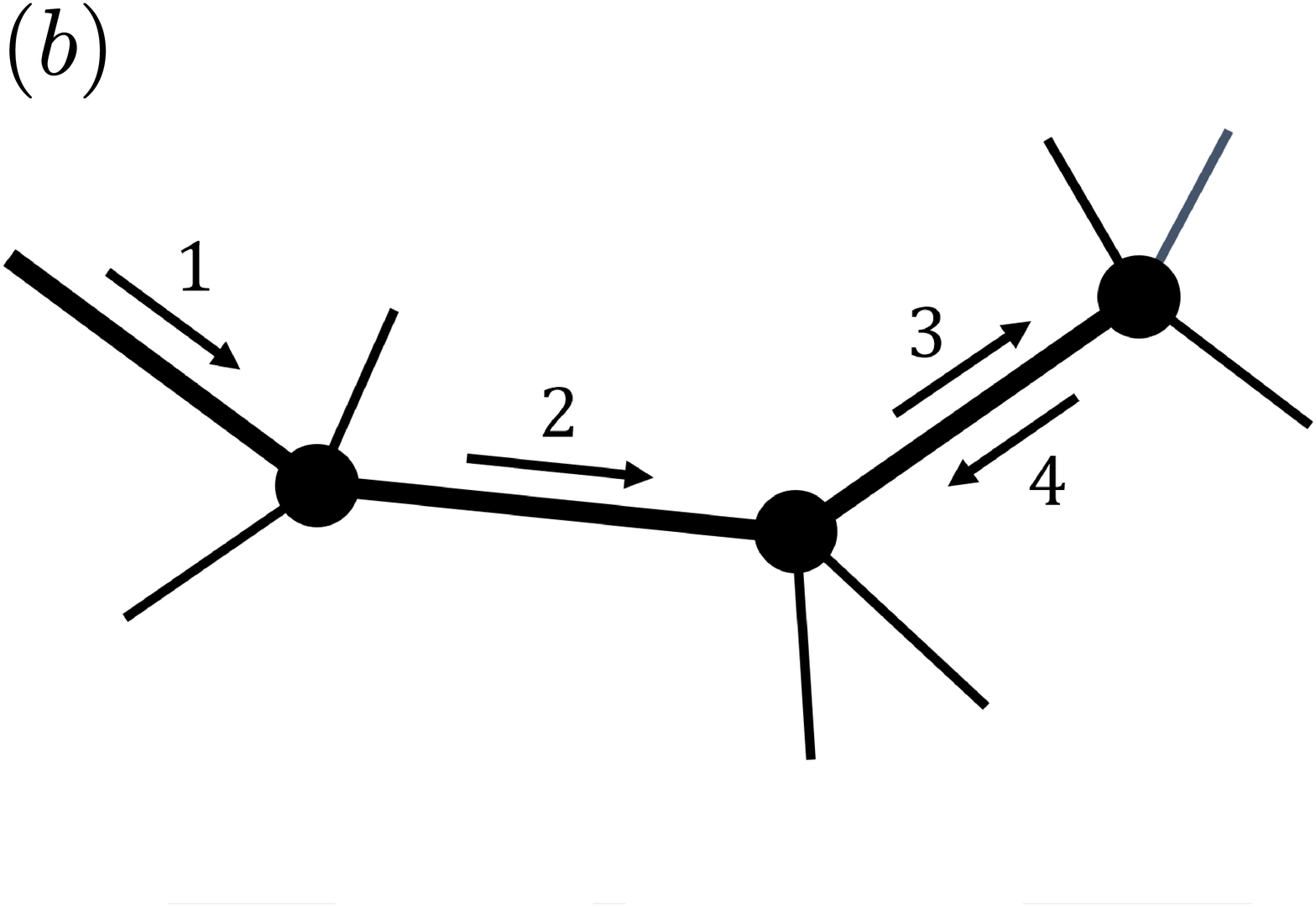}
}
\centerline{
\includegraphics[width=5.0cm]{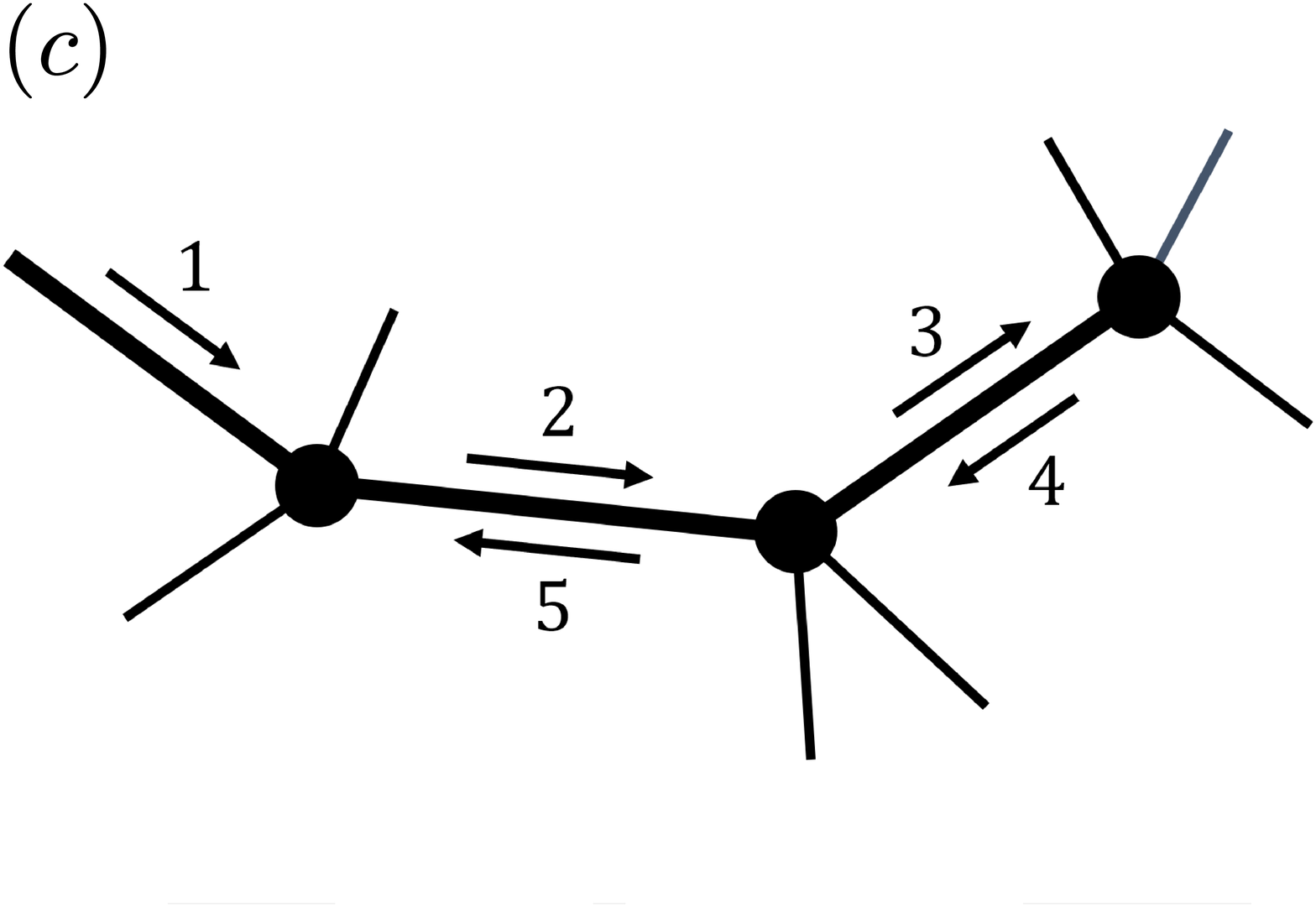}
\hspace{2cm}
%}
%\centerline{
\includegraphics[width=5.0cm]{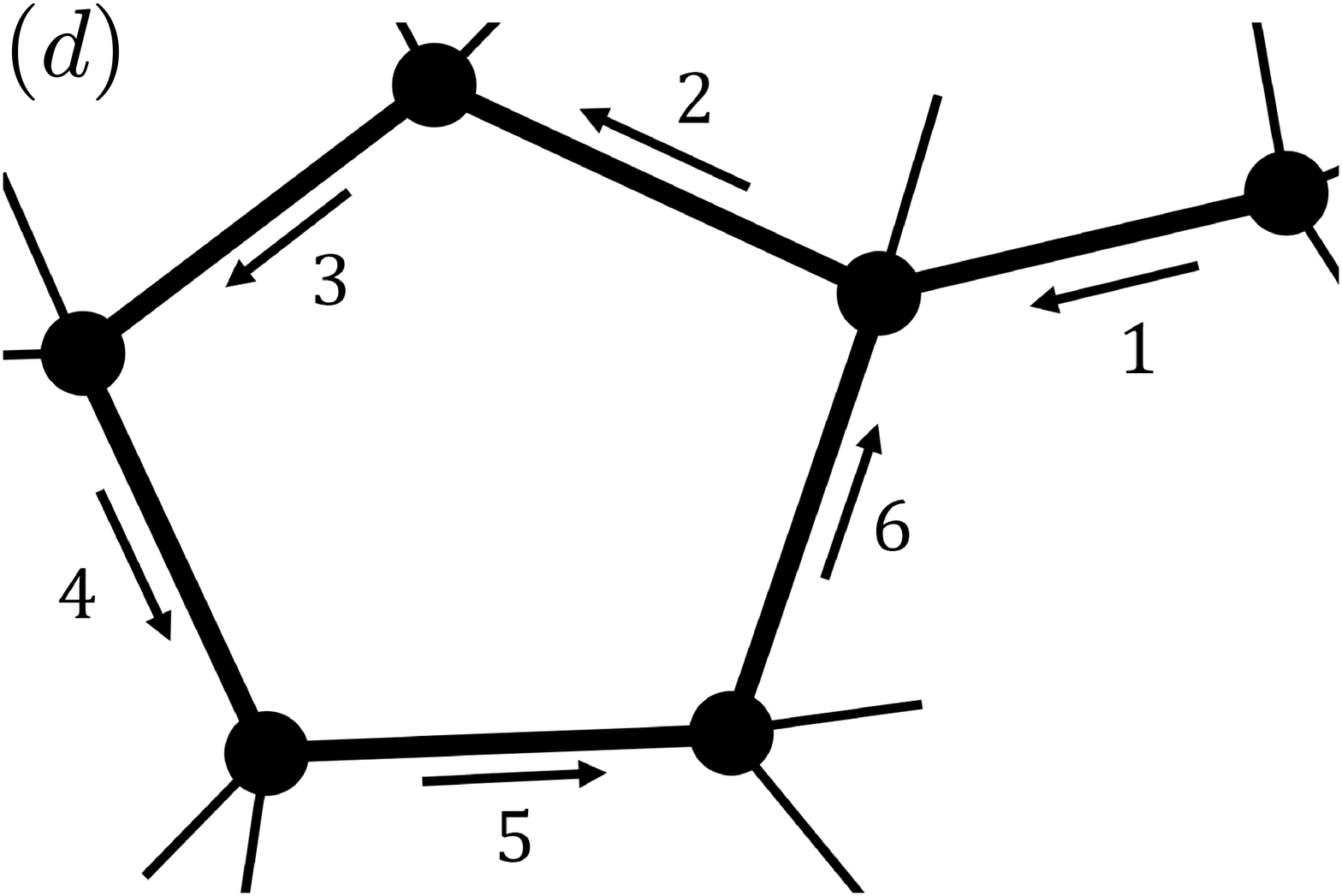}
}
\caption{
Schematic illustrations of possible events taking place along the path 
of an RW on an RRG:
(a) a path segment in which at each time step the RW enters a node
that has not been visited before; 
(b) a path segment that includes a backtracking step
into the previous node (step no. 4);
(c) a path segment that includes a backtracking step (step no. 4), which is 
followed by a retroceding step (step no. 5);
(d) a path that includes a retracing step (step no. 6) in which the RW hops into a node that was
visited a few time steps earlier. Retracing steps are not possible in the infinite
network limit and take place only in finite networks, which include cycles. 
Note that in this illustration the RRG
is of degree $c=4$.
}
\label{fig:1}
\end{figure}

In some of the time steps an RW may visit nodes that have not been visited before
while in other time steps it may revisit nodes that have already been visited before. 
For example, at each time step $t \ge 3$ the RW may backtrack into the previous node with 
probability of $1/c$.
In the infinite network limit the RRG exhibits a tree structure.
Therefore, in this limit the backtracking mechanism is the only way in which an RW
may hop from a newly visited node to a node that has already been visited before.
Such backtracking step may be followed by retroceding steps in which the RW
continues to go backwards along its own path.
However, in finite networks the RW may also utilize cycles to retrace its path 
and revisit nodes it has already visited three or more time steps earlier.
In Fig. \ref{fig:1} we present a
schematic illustration of some of the events that may take place along the path 
of an RW on an RRG.
In Fig. \ref{fig:1}(a) we show a path segment in which at each time step the RW enters a node
that has not been visited before.
In Fig. \ref{fig:1}(b) we show a path segment that includes a backtracking step,
in which the RW moves back into the previous node (step no. 4). 
In Fig. \ref{fig:1}(c) we show a path segment that includes a backtracking step
(step no. 4)
which is followed by a retroceding step (step no. 5).
In Fig. \ref{fig:1}(d) we show a path segment that includes a retracing step (step no. 6),
in which the RW enters a node that was visited five time steps earlier.

The mean number of distinct nodes 
that are visited by an RW on an RRG up to time $t$ 
is denoted by
$\langle S \rangle_t$.
The mean number of nodes in the complementary set of nodes that have not
been visited up to time $t$ is given by

\begin{equation}
\langle U \rangle_t = N - \langle S \rangle_t.
\label{eq:<U>_t}
\end{equation}

\noindent
The probability that an RW will step into a yet-unvisited node at time $t$
is given by 

\begin{equation}
\Delta_t = \langle S \rangle_{t} - \langle S \rangle_{t-1}.
\label{eq:Delta_t}
\end{equation}

\noindent
Using a generating function formulation based on the cavity method,
it was shown that in the infinite network limit
at sufficiently long times 
$\Delta_t \rightarrow \Delta$,
where
\cite{Debacco2015}

\begin{equation}
\Delta = \frac{c-2}{c-1}.
\label{eq:Delta_tc}
\end{equation}

\noindent
A similar result was obtained for RWs on
Bethe-Lattices 
\cite{Cassi1989,Martin2010}.
More precisely, $\Delta_t$ converges towards Eq. (\ref{eq:Delta_tc}) 
on a time scale of 
\cite{Tishby2021b}

\begin{equation}
\tau = \frac{2}{ \ln \left[ \frac{c^2}{4(c-1)} \right] }.
\end{equation}

\noindent
On a finite RRG of size $N$,
the probability of an RW that has already visited $s$ distinct
nodes to enter a yet-unvisited node in the next time step is given by
\cite{Tishby2021b}

\begin{equation}
\Delta(s) = \frac{c-2}{c-1} \left( 1 - \frac{s}{N} \right).
\label{eq:Deltas}
\end{equation}

\noindent
The complementary probability of an RW that has already visited $s$ distinct
nodes to enter a previously visited node in the next time step is given by

\begin{equation}
1 - \Delta(s) = \frac{1}{c-1} + \left( \frac{c-2}{c-1} \right) \frac{s}{N}.
\label{eq:1mDeltas}
\end{equation}

\noindent
The first term on the right hand side of Eq. (\ref{eq:1mDeltas}) accounts for the
probability of backtracking/retroceding 
[Figs. \ref{fig:1}(b) and \ref{fig:1}(c)],
while the second term accounts for the
probability of retracing
[Fig. \ref{fig:1}(d)].
The saturation term in Eq. (\ref{eq:Deltas}) is negligible at short times
and becomes dominant once the RW covers a large fraction of the network.
The analytical results for $\Delta(s)$ are presented in Fig. 4 of Ref. 
\cite{Tishby2021b}.
It is found to be in very good agreement with the results obtained 
from computer simulations
\cite{Tishby2021b}.
Note that Eq. (\ref{eq:Deltas}) is a slightly approximated version of the 
corresponding equation from Ref. \cite{Tishby2021b}.
In this approximation, we replaced $s-2$ by $s$ and $N-2$ by $N$.
The difference is negligible in the time scales considered in this paper
and helps to simplify the analysis.

\section{The master equation for $P_t(S=s)$}

Consider a trajectory of an RW on an RRG of size $N$.
There is no limit on the length of the trajectory and thus
the time $t$ may take values in the range $1 \le t < \infty$.
However, the number of distinct nodes $s$ visited by the RW
is bounded from above by the network size, namely $1 \le s \le N$.
The probability that an RW will visit $s$ distinct nodes
up to time $t$ is denoted by
$P_t(S=s)$.
Clearly, $P_t(S=s)=0$ for $s \ge t+1$.
Below we derive a master equation for the probability
$P_t(S=s)$.
To this end we 
utilize Eq. (\ref{eq:Deltas}), which provides the probability $\Delta(s)$ that an RW 
that has already visited $s$ distinct nodes 
will step into a yet-unvisited node in the next time step.
The complementary probability 
that the RW will step into a node that has already been visited before,
is given by $1-\Delta(s)$.
Note that the probability $P_{t+1}(S=s)$
is comprised of two contributions:
(a) the probability that the RW
has visited $s-1$ distinct nodes up to time $t$,
and that it subsequently entered a  
previously unvisited node at time $t+1$;
(b) the probability that the
RW has already visited $s$ distinct nodes up to time $t$,
and then entered a previously visited node at time $t+1$.
Taking into account these two contributions, 
we obtain 

\begin{equation}
P_{t+1}(S=s) =
\frac{c-2}{c-1}\left(1-\frac{s-1}{N}\right) P_{t}(S=s-1)
+
\left[ 1-\frac{c-2}{c-1}\left(1-\frac{s}{N}\right) \right]  P_{t}(S=s).
\label{eq:pnt0}
\end{equation}

\noindent
The time evolution of $P_t(S=s)$ can be expressed in terms of the forward difference

\begin{equation}
D_t P_t(S=s) = P_{t+1}(S=s) - P_t(S=s).
\label{eq:D_t}
\end{equation}

\noindent
Subtracting $P_{t}(S=s)$ from both sides of Eq. (\ref{eq:pnt0}), 
we obtain the discrete master equation
\cite{Gardiner2004,VanKampen2007,Krapivsky2010}

\begin{equation}
D_t P_t(S=s) =
\frac{c-2}{c-1} \left[  
\left(1-\frac{s-1}{N}\right) P_{t}(S=s-1)
-
\left(1-\frac{s}{N}\right)   P_{t}(S=s)
\right].
\label{eq:pnt0b}
\end{equation}

%\noindent
%Since the RW hops take place in discrete time steps,
%the replacement of the forward difference $D_t P_t(S=s)$ by a time
%derivative of the form $\partial P_t(S=s)/\partial t$ involves an approximation.
%In fact, it is closely related to the approximation made in the numerical integration
%of differential equations using the Euler method
%\cite{Butcher2003}.
%In the Euler method the time derivative $df_t/dt$ is replaced by $(f_{t+h}-f_t)/h$,
%where $h$ is a suitably chosen time step. 
%The case considered here corresponds to $h=1$.
%It can be shown that in such case the error is of the order of $1/N^2$.
%Replacing the difference on the left hand side 
%of Eq. (\ref{eq:pnt0b})
%by a time derivative,
%we obtain the master equation

%\begin{equation}
%\frac{\partial}{\partial t} P_t(S=s) =
%\frac{c-2}{c-1} \left[  
%\left(1-\frac{s-1}{N}\right) P_{t}(S=s-1) 
%-
%\left(1-\frac{s}{N}\right)   P_{t}(S=s)
%\right].
%\label{eq:pnt0c}
%\end{equation}

\noindent
The master equation consists of a set of coupled difference equations
for $P_t(S=s)$, $s=2,\dots N$, at $t \ge 2$.
% or in other words it is a partial difference 
%equation. 
%Such master equations are often solved by direct numerical integration
%using Runge-Kutta or other methods
%\cite{Press1992}.
The initial condition is given by
$P_{t=2}(S=s)=\delta_{s,2}$.
Note also that in the first time step
$P_{t=1}(S=s)=\delta_{s,1}$,
while at $t \ge 2$ the probability
$P_{t}(S=1)=0$.
A special property of Eq. (\ref{eq:pnt0b}) 
%and (\ref{eq:pnt0c}) 
is that probability flows only upwards along
the $s$ axis from $s-1$ to $s$. 
This means that Eq. (\ref{eq:pnt0b}) 
%and (\ref{eq:pnt0c}) 
does not support a steady-state solution, apart
from the absorbing state solution imposed by the finite size of the network,
which is given by $P(S=s)=\delta_{s,N}$.
In general, an absorbing state is a state which once entered cannot be left.
The state $s=N$ is the only absorbing state of the RW.
Moreover, every single trajectory will eventually reach this absorbing state.
Thus, the Markov chain describing the covering process of an RRG by an RW
is referred to as an absorbing chain
\cite{Kemeny1960}.

\section{The solution of the master equation for $N \rightarrow \infty$}

In the process of solving the master equation, it is instructive to first
consider the infinite network limit.
In this limit the discrete master equation (\ref{eq:pnt0b})
is reduced to

\begin{equation}
D_t P_t(S=s) =
\left( \frac{c-2}{c-1} \right)
\left[     
P_{t}(S=s-1)
-
P_{t}(S=s) 
\right].
\label{eq:pnt0cinf}
\end{equation}

\noindent
Unlike the case of a finite network in which the master equation consists of $N-1$
equations for $s=2,3,\dots,N$, in the infinite network limit the master equation 
consists of an infinite number of equations, for $s \ge 2$.
It is easy to verify that the
solution of Eq. (\ref{eq:pnt0cinf}) for $t \ge 2$ is given by
a binomial distribution of the form

\begin{equation}
P_t(S=s) = \binom{t-2}{s-2} \left( \frac{c-2}{c-1} \right)^{s-2} \left( \frac{1}{c-1} \right)^{t-s},
\label{eq:Ptsinf}
\end{equation}

\noindent
where $\binom{t}{s}$ is the binomial coefficient.
 
The mean number of distinct nodes visited by an RW
up to time $t$ is given by

\begin{equation}
\langle S \rangle_t = 2 + \frac{c-2}{c-1} (t-2),
\end{equation}

\noindent
while the variance of $P_t(S=s)$ is given by

\begin{equation}
{\rm Var}_t(S) = \frac{ c-2 }{ (c-1)^2 } (t-2).
\end{equation}

In the next section we present the results for $P_t(S=s)$ in finite networks.
In the limit of $N \rightarrow \infty$
these results must converge towards Eq. (\ref{eq:Ptsinf}).
In order to show this,
it is useful to express Eq. (\ref{eq:Ptsinf}) in a different form. 
Inserting the Binomial expansion

\begin{equation}
\left( \frac{1}{c-1} \right)^{t-s} =
\left( 1 - \frac{c-2}{c-1} \right)^{t-s} =
\sum_{m=0}^{t-s}  (-1)^m   \binom{t-s}{m}  \left( \frac{c-2}{c-1} \right)^m,
\end{equation}

\noindent
into Eq. (\ref{eq:Ptsinf}), we obtain

\begin{equation}
P_t(S=s) = 
\binom{t-2}{s-2} \sum_{m=0}^{t-s}  (-1)^m   \binom{t-s}{m}  \left( \frac{c-2}{c-1} \right)^{m+s-2}.
\label{eq:Ptsbinom2}
\end{equation}

\noindent
In the next section we indeed show that Eq. (\ref{eq:Ptsbinom2}) is obtained
as the $N \rightarrow \infty$ limit of the distribution $P_t(S=s)$ in finite networks.

\section{The solution of the master equation for finite networks}

In Appendix A we use a generating function approach to solve the discrete master equation
[Eq. (\ref{eq:pnt0b})] for the case of a finite network that consists of $N$ nodes. 
The solution is given by

\begin{eqnarray}
P_t(S=s) &=& 
\sum_{v=s-2}^{t-2} 
(-1)^{v-s} 
\binom{t-2}{v} 
\left[ \left( \frac{c-2}{c-1} \right) \frac{1}{N} \right]^{v}
\times
\nonumber \\
\ \ \ \ \ \ \ \ \  & &
\sum_{m=s-2}^{ \min{ \{ v,N-2 \} }} 
m!
\Big\{
\begin{array}{l}
v  \\
m       
\end{array}
\Big\}
\binom{N-2}{m}
\binom{m}{s-2},
\label{eq:P_t(S=s)!s}
\end{eqnarray}

\noindent
where 

\begin{equation}
\Big\{
\begin{array}{l}
v  \\
m       
\end{array}
\Big\} =
\frac{1}{m!} \sum_{k=0}^{m} (-1)^k \binom{m}{k} (m-k)^v
\end{equation}

\noindent
is the Stirling number of the second kind
\cite{Olver2010}.
The Stirling number of the second kind [which is also denoted by $S(v,m)$]
represents the number of ways to partition a set of $v$ labeled objects into $m$
non-empty subsets.
Clearly, this solution presented by Eq. (\ref{eq:P_t(S=s)!s}) satisfies the condition that
$P_t(S=s)=0$ for $s \ge t+1$.

In the limit of $N \rightarrow \infty$,
the solution for $P_t(S=s)$ on a finite network, given by
Eq. (\ref{eq:P_t(S=s)!s}), is reduced to the solution 
on an infinite network, given by 
Eq. (\ref{eq:Ptsbinom2}).
To show this property we 
expand the right hand side of Eq. (\ref{eq:P_t(S=s)!s})
in powers of $1/N$, under the condition that $t < N$. 
The zero-order term of this expansion is obtained from the
$m=v$ term in the second sum.
Replacing the second sum by this term alone,
the resulting expression is
found to
be identical to Eq. (\ref{eq:Ptsbinom2}). 
Thus, in the infinite system limit the solution of Eq. (\ref{eq:pnt0b}),
which describes the time evolution of $P_t(S=s)$ on finite networks is reduced to the
solution of Eq. (\ref{eq:pnt0cinf}) that describes the infinite system limit.

Below we derive an alternative expression for 
$P_t(S=s)$.
Exchanging the order of the summations in Eq. (\ref{eq:P_t(S=s)!s})
and rearranging terms, we obtain

\begin{eqnarray}
P_t(S=s) &=& 
\sum_{m=s-2}^{ N-2 } 
(-1)^{m-s}
\binom{N-2}{m}
\binom{m}{s-2}
\times
\nonumber \\
& &
(-1)^m m! 
\sum_{v=0}^{t-2} \binom{t-2}{v} 
\Big\{
\begin{array}{l}
v  \\
m       
\end{array}
\Big\}
\left[ - \left( \frac{c-2}{ c-1 } \right) \frac{1}{N} \right]^{v}.
\label{eq:P_t(S=s)!s2}
\end{eqnarray}

\noindent
Note the lower limit of the second sum in Eq. (\ref{eq:P_t(S=s)!s2}) is 
$0$ rather than $m$. This is due to the fact that for $v<m$ the Stirling
number satisfies 
$\Big\{
\begin{array}{l}
v  \\
m       
\end{array}
\Big\}
= 0$
Using identity (7.7) in Ref. \cite{Boya2018},
which is given by

\begin{equation}
(-1)^m m! \sum_{v=0}^n \binom{n}{v} 
\Big\{
\begin{array}{l}
v  \\
m       
\end{array}
\Big\}
z^v
= 
\sum_{k=0}^m 
(-1)^k \binom{m}{k}  (1+zk)^n,
\end{equation}

\noindent
we obtain

\begin{equation}
P_t(S=s) =
\sum_{m=s-2}^{N-2} (-1)^{m-s} \binom{N-2}{m} \binom{m}{s-2}
\sum_{k=0}^m (-1)^k  \binom{m}{k}  
\left[ 1 - \left( \frac{c-2}{c-1} \right) \frac{k}{N}   \right]^{t-2}.
\label{eq:PtSalt}
\end{equation}

The right hand sides of
Eqs. (\ref{eq:P_t(S=s)!s}) and (\ref{eq:PtSalt})
include double sums.
It is thus useful to consider the number of terms included in
these sums, in order to compare the computational effort
involved in the calculation of $P_t(S=s)$ using
Eqs. (\ref{eq:P_t(S=s)!s}) and (\ref{eq:PtSalt}).
In Eq. (\ref{eq:P_t(S=s)!s}),
for $t < N$
the number of terms in the double sum  
scales like $t^2$ 
(assuming that 
$\Big\{
\begin{array}{l}
v  \\
m       
\end{array}
\Big\}$
is evaluated using a lookup table).
For $t > N$ the number of terms
scales like $Nt$.
As a result, for $t>N$ it becomes difficult to evaluate the right
hand side of Eq. (\ref{eq:P_t(S=s)!s}).
The number of terms in the double sum of Eq. (\ref{eq:PtSalt})
scales like $N^2$ and does not depend on the time $t$.
As a result, at long times $t > N$ the evaluation of $P_t(S=s)$ using 
Eq. (\ref{eq:PtSalt}) 
is more efficient than  
Eq. (\ref{eq:P_t(S=s)!s}).
In light of these considerations, the evaluation of $P_t(S=s)$ 
is done using Eq. (\ref{eq:P_t(S=s)!s}) for $t \le N$ and using Eq. (\ref{eq:PtSalt})
for $t >N$.
It is important to emphasize that regardless of the
efficiency considerations discussed above, for any value of $t$
Eqs. (\ref{eq:P_t(S=s)!s}) and (\ref{eq:PtSalt}) are equivalent.

In the long time limit, where $t \gg N$, one can use the approximation

\begin{equation}
\left[ 1 - \left( \frac{c-2}{c-1} \right) \frac{k}{N} \right]^{t-2}
\simeq \exp \left[  - \left( \frac{c-2}{c-1} \right) \frac{k}{N} (t-2) \right].
\label{eq:expo}
\end{equation}

\noindent
Inserting the right hand side of Eq. (\ref{eq:expo})
into Eq. (\ref{eq:PtSalt}) and carrying out the summation
over $k$, we obtain

\begin{equation}
P_t(S=s) \simeq
\sum_{m=s-2}^{N-2} (-1)^{m-s} \binom{N-2}{m} \binom{m}{s-2}
\left[ 1 - e^{ - \left( \frac{c-2}{c-1} \right) \frac{t-2}{N}  }   \right]^m.
\label{eq:PtSalt2}
\end{equation}

\noindent
Writing the binomial coefficients explicitly in terms of the factorials,
Eq. (\ref{eq:PtSalt2}) becomes

\begin{eqnarray}
P_t(S=s) &\simeq&
\frac{(N-2)!}{(s-2)!}
\sum_{m=s-2}^{N-2} 
\frac{ (-1)^{m-s} }{ (m-s+2)! (N-2-m)! }
\times
\nonumber \\
& & \left[ 1 - e^{ - \left( \frac{c-2}{c-1} \right) \frac{t-2}{N}  }   \right]^m.
\label{eq:PtSalt5}
\end{eqnarray}

\noindent
Shifting the summation index from $m$ to $r=m-(s-2)$,
we obtain

\begin{equation}
P_t(S=s) \simeq
\frac{(N-2)!}{(s-2)!}
\sum_{r=0}^{N-s} 
\frac{ (-1)^{r} }{ r! (N-s-r)! }
\left[ 1 - e^{ - \left( \frac{c-2}{c-1} \right) \frac{t-2}{N}  }   \right]^{r+s-2}.
\label{eq:PtSalt6}
\end{equation}

\noindent
Multiplying the numerator and the denominator by $(N-s)!$
and carrying out the summation over $r$, we obtain

\begin{eqnarray}
P_t(S=s) &\simeq&
\binom{N-2}{s-2}
\left[ 1 - e^{ - \left( \frac{c-2}{c-1} \right) \frac{t-2}{N}  }  \right]^{s-2} 
\left[   e^{ - \left( \frac{c-2}{c-1} \right) \frac{t-2}{N}  }   \right]^{N-s}.
\label{eq:PtSalt7}
\end{eqnarray}

\noindent
This implies that $P_t(S=s)$ can be expressed as a binomial distribution.
The term in the first square bracket represents the probability that a random
node has already been visited by the RW up to time $t$ while the term in the
second square bracket represents the probability that a random node has not
yet been visited up to time $t$.

The tail distribution $P_t(S>s)$ is given by

\begin{equation}
P_t(S>s) = \sum_{s'=s+1}^{N} P_t(S=s').
\label{eq:PtStail}
\end{equation}

\begin{figure}
\centerline{
\includegraphics[width=15cm]{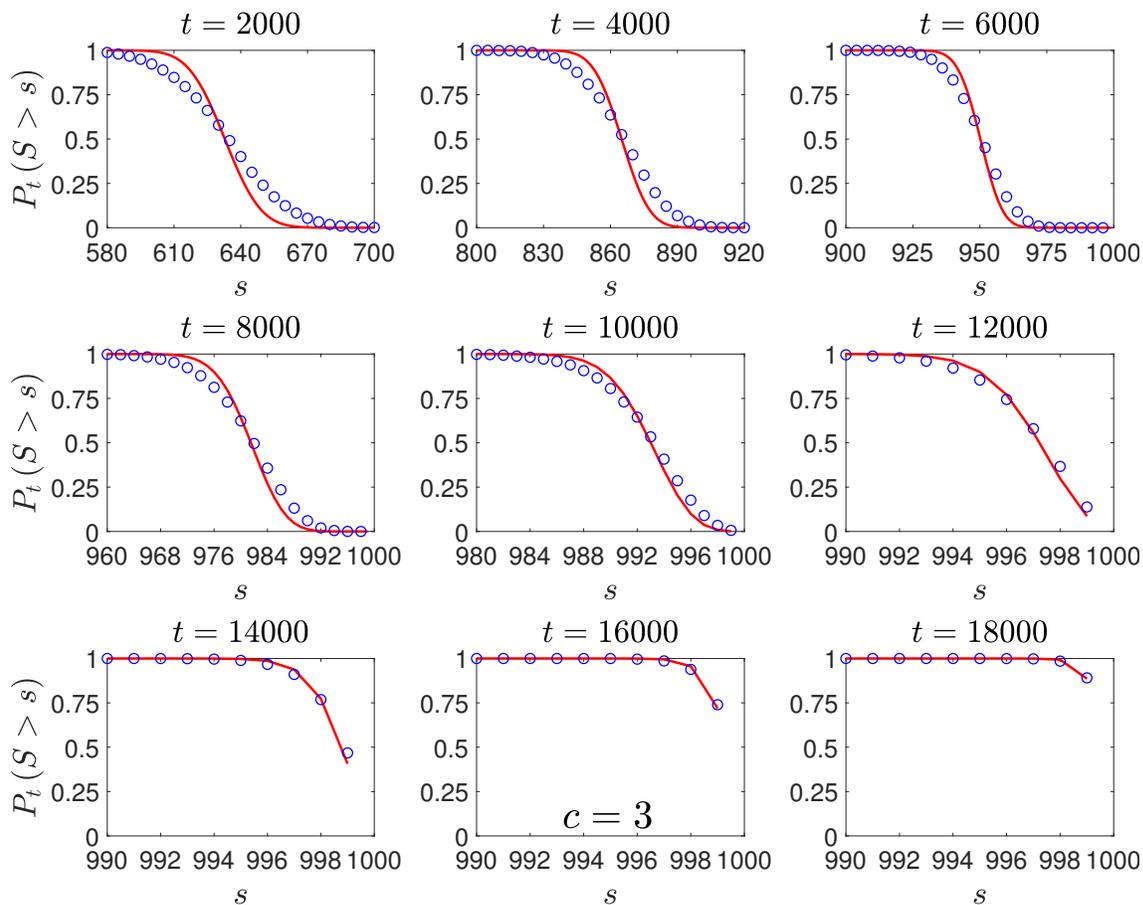}
}
\caption{
Analytical results (solid lines) for the
probability $P_t(S>s)$ that an RW will visit at least $s$ distinct nodes
up to time $t$ on a random regular graph of size
$N=1000$ and node degree $c=3$. 
The results are presented for nine values of the time $t$ 
from $t=2000$ to $t=18000$.
At early times the
analytical results, obtained from Eq. (\ref{eq:PtStail}),
deviate significantly from the results obtained
from computer simulations (circles).
As time proceeds the agreement between the analytical  
and the simulation results gradually improves and becomes
very good beyond $t=10000$.
}
\label{fig:2}
\end{figure}

In Fig. \ref{fig:2} we present the tail distribution  
$P_t(S>s)$ for a network of size $N=1000$ with degree $c=3$ 
and nine values of $t$ from $t=2000$ to $t=18000$.
At early times the
analytical results (solid lines), obtained from Eq. (\ref{eq:PtStail}),
deviate significantly from the results obtained
from computer simulations (circles).
As time proceeds the agreement between the analytical  
and the simulation results gradually improves and becomes
very good beyond $t=10000$.

The discrepancy between the analytical and the simulation results 
in Fig. \ref{fig:2} are due to short range temporal correlations
in the RW trajectories for RRGs of low degree and are most
pronounced at $c=3$.
To explain this point consider an RW on an RRG of low degree $c$
that has visited $s$ distinct nodes
up to time $t-1$ and 
at time $t$ it steps into a previously visited node $x_t$
(such that the number of distinct nodes visited remains $s$).
Under these conditions,
the probability that at time $t+1$ the RW will also step into a previously
visited node is higher than mean-field value of $1-\Delta(s)$,
where $\Delta(s)$ is given by Eq. (\ref{eq:Deltas}).
This is due to the fact that if $x_t$ has been visited before, it 
must have entered $x_t$ via one of its neighbors and 
must have left $x_t$ via the same neighbor or another neighbor.
In the case of a small degree $c$, the one or two neighbors
visited before and after the previous visit of $x_t$ represent a large fraction
of all the neighbors of $x_t$.
As a result, the probability that the RW will step into a previously visited
node at time $t+1$ is higher than the mean-field result.
Similarly, in case that the node $x_t$ visited at time $t$ has not been
visited before, the probability that at time $t+1$ the RW will step into
a previously visited node is lower than the mean-field result.
Due to these correlations, in RRGs of a low degree $c$ 
there may be instances of the RW trajectory which include long streaks of
steps in which the RW revisits previously visited nodes.
Similarly, there may be instances of the RW trajectory which include long
streaks of steps in which the RW visits new, yet-unvisited nodes.
These correlations thus broaden the distribution $P_t(S=s)$
obtained from computer simulations, compared to the analytical results.
While the temporal correlations are short ranged, their effect 
on the number of distinct nodes visited up to time $t$ 
may accumulate, giving rise to a non-diminishing long term effect.
From Eq. (\ref{eq:TC1}) one concludes that by the time the whole
network is covered, each node has been visited on average
$\sim \ln N$ times.
This implies that for $c>\ln N$ the correlations discussed above
are negligible and the results for $P_t(S=s)$, obtained from
the master equation, are probably exact.
In fact, these results remain highly accurate even for much smaller values of $c$.
Moreover, noticeable discrepancies in $P_t(S=s)$ are observed only for the smallest 
possible values of $c$, namely $c=3$ and $4$.

\begin{figure}
\centerline{
\includegraphics[width=15cm]{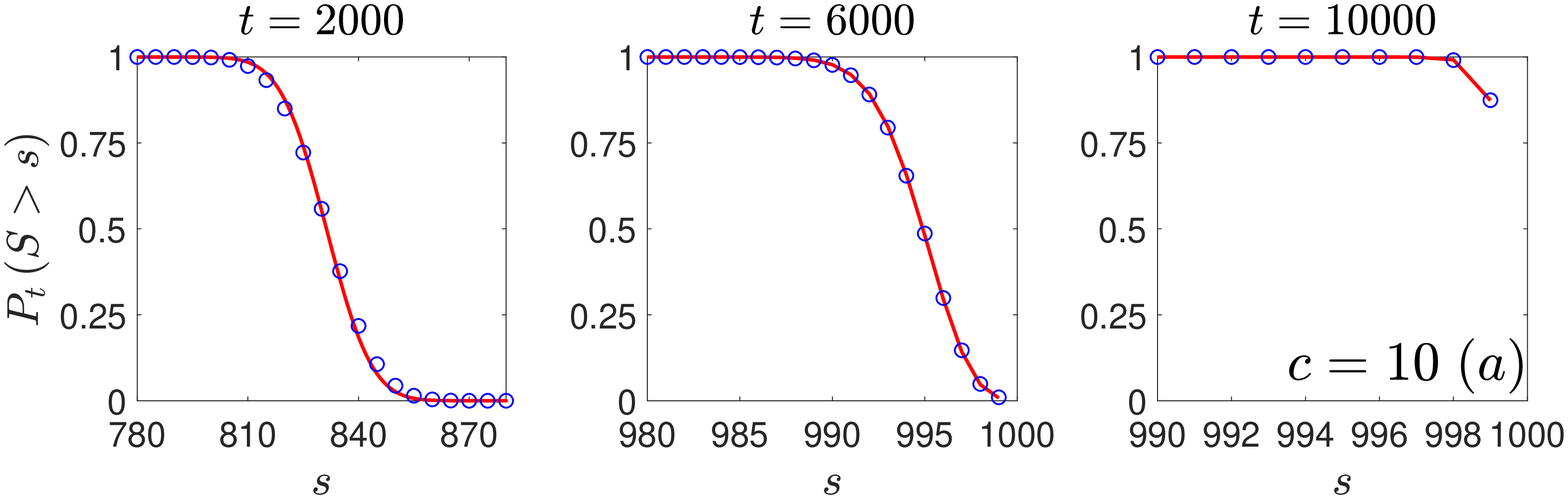}
}
\vspace{0.2in}
\centerline{
\includegraphics[width=15cm]{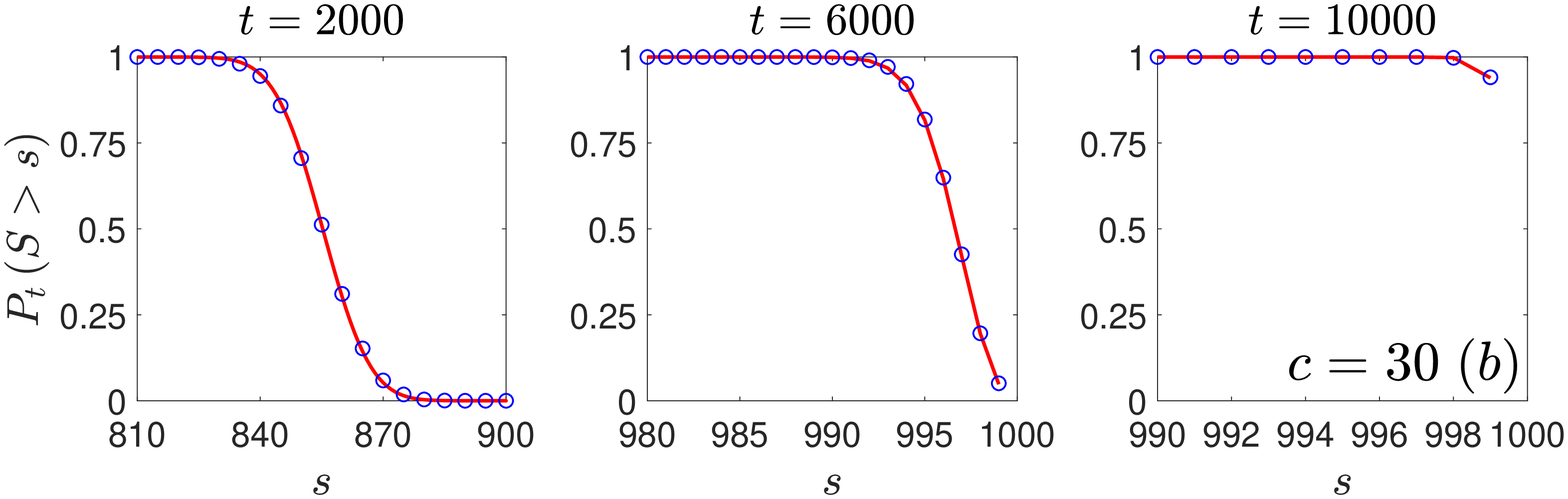}
}
\caption{
The probability $P_t(S>s)$ that an RW will visit at least $s$ distinct nodes
up to time $t$ on a random regular graph of size
$N=1000$ and node degree $c=10$ (a) and $c=30$ (b). 
The results obtained for three values of the time $t$ are shown: 
$t=2000$ (left), $t=6000$ (center), $t=10000$ (right).
The analytical results (solid lines), 
obtained from Eq. (\ref{eq:PtStail})
are in very good agreement with the results obtained
from numerical simulations (circles).
}
\label{fig:3}
\end{figure}

In Fig. \ref{fig:3} we present the tail distribution  
$P_t(S>s)$ for a network of size $N=1000$ with degrees $c=10$ (a) and $c=30$ (b)
at times $t=2000$ (left), $t=6000$ (center) and $t=10000$ (right).
The analytical results are 
in very good agreement with the results obtained from computer simulations.
As time evolves the sigmoid curve in $P_t(S>s)$ 
slides to the right and becomes narrower 
as it approaches the boundary at $s=N$.
At late times the probability $P_t(S>N-1)$,
which is the probability that the RW has completed covering
the whole network up to time $t$,
becomes nonzero and continues to increase as time evolves.

For the simulations we generated a large number (typically 100) 
of random instances of the RRG 
consisting of $N$ nodes of degree $c$, using the procedure 
presented in Sec. 2. For each network instance, we generated
a large number (typically 100) of RW trajectories, where each trajectory starts 
from a random initial node $i=x_1$ at time $t=1$. 
The simulation results are obtained by averaging over all these trajectories.
In the simulations, at each time step $t$ the RW selects randomly  
one of the $c$ neighbors of the node $x_{t-1}$,
where the probability of each neighbor to be selected 
is $1/c$. It then hops to the selected node, denoted by $x_t$.
The number of distinct nodes visited up to time $t$ in a given RW trajectory
is denoted by $s_t$. 
Each RW trajectory is terminated once it covers all the nodes in the network,
namely when $s_t=N$.
The cover time is thus equal to the length of the trajectory.
The trajectory $x_0,x_1,\dots,x_t$ is recorded for further analysis.

Another interesting probability is the 
inverse of the distribution $P_t(S=s)$, namely the conditional
probability 
$P(T=t|s)$
that an RW has pursued $t$ time steps,
given that it has visited $s$ distinct nodes.
This probability can be obtained by marginalizing 
$P_t(S=s)$, namely

\begin{equation}
P(T=t|s) = \frac{ P_t(S=s) }{ \sum_{t=1}^{\infty}   P_t(S=s) }.
\label{eq:PTtsi}
\end{equation}
 
\noindent
Clearly, $P(T=1|1)=1$.
The probability $P(T=t|s)$ is defined in the range of 
$2 \le s \le N-1$ and $t \ge s$. 
In Appendix B we use a generating function approach to obtain
a closed-form expression for $P(T=t|s)$ 
[Eq. (\ref{eq:PTts})] and to calculate its
mean [Eq. (\ref{eq:ETs1})] and variance [Eq. (\ref{eq:VTsb})].

\begin{figure}
\centerline{
\includegraphics[width=15cm]{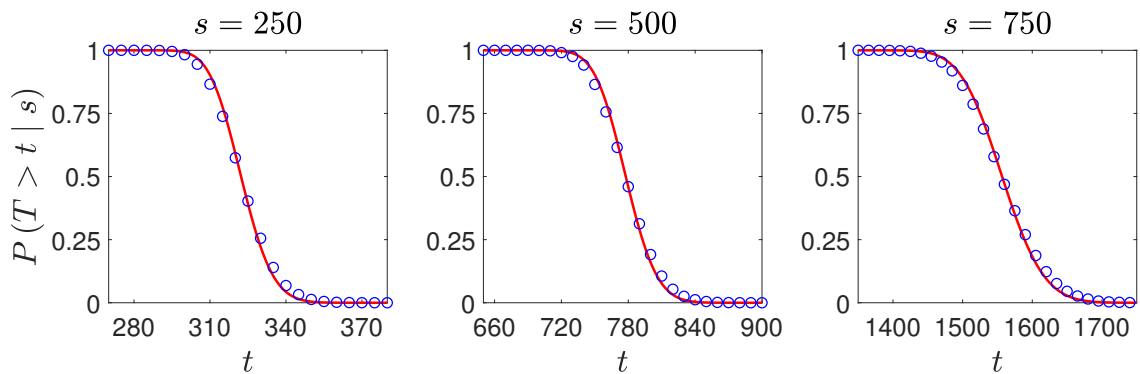}
}
\caption{
The conditional tail distribution $P(T>t|s)$  
on the elapsed time given that the RW has visited $s$ distinct nodes,
for an RRG of size
$N=1000$, degree $c=10$ and
$s=N/4$ (left) , $s=N/2$ (center) and $s=3N/4$ (right).
The analytical results (solid lines), 
obtained from
Eqs. (\ref{eq:PTts}) and (\ref{eq:PTts7})  are in very good agreement with the 
results obtained from computer simulations (circles).
}
\label{fig:4}
\end{figure}

In Fig. \ref{fig:4} we present the tail distribution  
$P(T>t|s)$ for a network of size $N=1000$ with degree $c=10$ 
and $s=N/4$ (left), $s=N/2$ (middle) and $s=3N/4$ (right).
The analytical results are 
in very good agreement with the results obtained from computer simulations.
It is found that as $s$ is increased the sigmoid-like function shifts to the 
right and broadens.

\section{The mean and variance of $P_t(S=s)$}

In Appendix C we use a generating function formulation to calculate the moments
of $P_t(S=s)$. 
The first moment is given by

\begin{equation}
\langle S \rangle_t = 2 
+ (N-2) \left[ 1 - e^{ -  \left( \frac{c-2}{c-1} \right) \frac{t-2}{N}   } \right],
\label{eq:Stlate2}
\end{equation}

\noindent
which coincides with previous results  
obtained using other methods
\cite{Tishby2021b}.
Thus, the mean number of nodes in the complementary set of nodes that have not
been visited by the RW up to time $t$ is given by

\begin{equation}
\langle U \rangle_t =   
(N-2) e^{ -  \left( \frac{c-2}{c-1} \right) \frac{t-2}{N}   }.
\label{eq:Utlate2}
\end{equation}

\begin{figure}
\centerline{
\includegraphics[width=8cm]{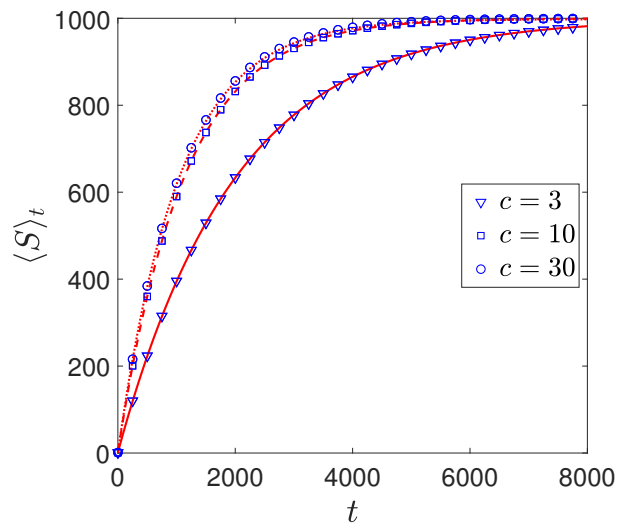}
}
\caption{
Analytical results for
the mean number $\langle S \rangle_t$ 
of nodes visited by an RW up to time $t$ 
on an RRG of size
$N=1000$
and degrees 
$c=3$ (solid line), $c=10$ (dashed line)
and $c=30$ (dotted line).
The analytical results,
obtained from Eq. (\ref{eq:Stlate2}),
are in very good agreement with the results
obtained from computer simulations (triangles, squares and circles, respectively).
}
\label{fig:5}
\end{figure}

In Fig. \ref{fig:5} 
we present analytical results for
the mean number $\langle S \rangle_t$ 
of nodes visited by an RW up to time $t$
on an RRG of size
$N=1000$
and degrees 
$c=3$ (solid line), $c=10$ (dashed line)
and $c=30$ (dotted line).
The analytical results,
obtained from Eq. (\ref{eq:Stlate2}),
are in very good agreement with the results
obtained from computer simulations (circles).

The variance of $P_t(S=s)$ is given by

\begin{equation}
{\rm Var}_t(S) = \langle S^2 \rangle_t - (\langle S \rangle_t)^2.
\label{eq:VAR}
\end{equation}

\noindent
In Appendix C we  
obtain the second moment $\langle S^2 \rangle_t$.
Inserting the results for the first and second moments into Eq. (\ref{eq:VAR}), 
we obtain  

\begin{eqnarray}
{\rm Var}_t(S) &=&
(N-2) \left[ 1 - \left( \frac{c-2}{c-1} \right) \frac{1}{N} \right]^{t-2}
\nonumber \\
&-& (N-2)^2 \left[ 1 - \left( \frac{c-2}{c-1} \right) \frac{1}{N} \right]^{2(t-2)}
\nonumber \\
&+& (N-3)(N-2) \left[ 1 - 2 \left( \frac{c-2}{c-1} \right) \frac{1}{N} \right]^{t-2}.
\label{eq:VartS}
\end{eqnarray}

\noindent
This is a new result, which could not be obtained using the
methods of Ref. \cite{Tishby2021b}.

\begin{figure}
\centerline{
\includegraphics[width=15cm]{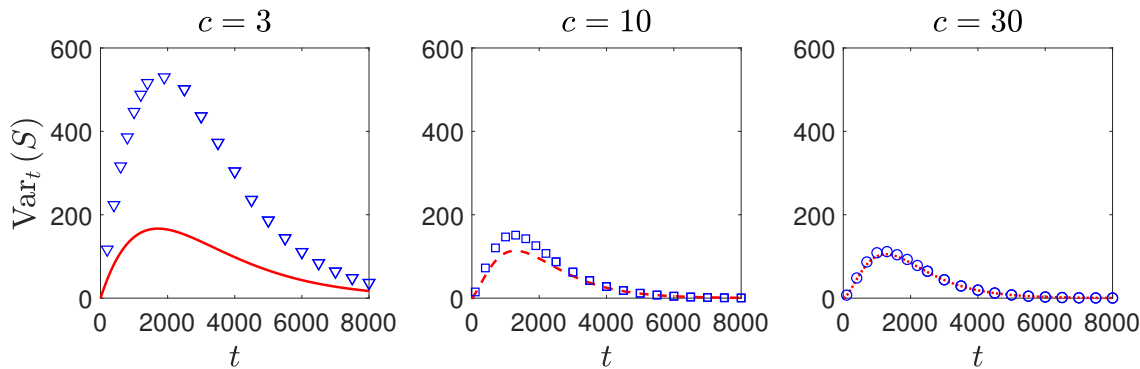}
}
\caption{
Analytical results for
the variance ${\rm Var}_t(S)$ 
of the distribution $P_t(S=s)$ 
of the number of distinct nodes
visited up to time $t$ 
for an RW
on an RRG of size
$N=1000$
and degree  
$c=3$ (solid line, left), $c=10$ (dashed line, center)
and $c=30$ (dotted line, right),
and the corresponding simulation results 
(triangles, squares and circles, respectively).
For $c=3$ the results for the variance obtained from computer simulations 
(triangles) are much larger than those obtained from the analytical calculations.
This discrepancy is due to short range temporal correlations in the RW trajectories
that are most pronounced at small values of $c$.
For $c=10$ the agreement between the analytical results (dashed line) and the simulation
results (squares) is significantly improved, while for $c=30$ there is a good agreement.
}
\label{fig:6}
\end{figure}

In Fig. \ref{fig:6} 
we present analytical results for
the variance ${\rm Var}_t(S)$ 
as a function of $t$ for an RW
on an RRG of size
$N=1000$
and degrees 
$c=3$ (solid line, left), $c=10$ (dashed line, center)
and $c=30$ (dotted line, right).
For $c=3$ the simulation results (triangles)
exhibit a significant deviation from the analytical results (solid line),
obtained from Eq. (\ref{eq:VartS}).
This deviation is due to temporal correlations between the probabilities
to visit previously unvisited nodes in successive time steps,
which are most pronounced for small values of $c$.
However, note that at the peak of ${\rm Var}_t(S)$ for $c=3$
in Fig. \ref{fig:6}, obtained for $t \simeq 2000$, the standard deviation obtained from
the simulations is $\sigma_t(S) = \sqrt{ {\rm Var_t(S)} } \simeq 24$,
compared to $\sigma_t(S) \simeq 14$ obtained from the theoretical calculations.
Both values are small compared to $\langle S \rangle_t \simeq 600$ at 
$t \simeq 2000$, namely $P_t(S=s)$ exhibits a narrow peak around $\langle S \rangle_t$.
For $c=10$ the
analytical results (dashed line)
are in better agreement with the results
obtained from computer simulations (squares),
while for $c=30$ there is a good agreement,
showing that the mean-field argument becomes more accurate as $c$
is increased.

\section{The distribution of cover times}

Inserting $s=N$ in the distribution $P_t(S=s)$ one obtains
$P_t(S=N)$, 
which is the probability that the
RW has visited all the nodes in the network up to time $t$.
In fact, this coincides with the cumulative probability 
of the cover time, namely

\begin{equation}
P(T_{\rm C} \le t) = P_t(S=N).
\label{eq:PTCtPtsN}
\end{equation}

\noindent
The tail distribution of cover times is given by

\begin{equation}
P(T_{\rm C} > t) = 1 - P(T_{\rm C} \le t).
\end{equation}

\noindent
Therefore,

\begin{equation}
P(T_{\rm C} > t) = 1 - P_t(S=N).
\label{eq:PTCgt}
\end{equation}

\noindent
Inserting $s=N$ in Eq. (\ref{eq:P_t(S=s)!s})
and plugging in the right hand side into Eq. (\ref{eq:PTCgt}),
we obtain
 
\begin{equation}
P(T_{\rm C} > t) = 
1 - (N-2)!  \sum_{v=N-2}^{t-2} 
(-1)^{v-N}
\binom{t-2}{v} 
\Big\{
\begin{array}{c}
v  \\
N-2       
\end{array}
\Big\}
\left[ \left( \frac{c-2}{c-1} \right) \frac{1}{N} \right]^{v}
\label{eq:P_t(S=N)!2}.
\end{equation}
 
\noindent
The number of terms in the sum on the right hand side of Eq. (\ref{eq:P_t(S=N)!2})
scales like $t-N$. It is thus efficient as long as $t-N$ is not too large.
For longer times it is more efficient to extract the distribution of cover times from
Eq. (\ref{eq:PtSalt}).
Inserting $s=N$ in Eq. (\ref{eq:PtSalt}), we obtain

\begin{equation}
P(T_{\rm C} > t) = 1 - \sum_{k=0}^{N-2} (-1)^k \binom{N-2}{k} 
\left[ 1 - \left( \frac{c-2}{c-1} \right) \frac{k}{N} \right]^{t-2}.
\end{equation}

In the long time limit $t \gg N$, 
one could use Eq. (\ref{eq:PtSalt7}) to approximate the cover time.
Inserting $s=N$ in Eq. (\ref{eq:PtSalt7}), we obtain

\begin{equation}
P(T_{\rm C} > t) \simeq
1 - \left[ 1 - e^{- \left( \frac{c-2}{c-1} \right) \frac{t-2}{N} } \right]^{N-2}.
\label{eq:Gumbel1}
\end{equation}

\noindent
Moreover, Eq. (\ref{eq:Gumbel1}) can be approximated by

\begin{equation}
P(T_{\rm C} > t) \simeq
1 - \exp \left[ - (N-2) e^{- \left( \frac{c-2}{c-1} \right) \frac{t-2}{N} } \right].
\label{eq:Gumbel2}
\end{equation}

\noindent
Interestingly, Eq. (\ref{eq:Gumbel2}) can be written in the form

\begin{equation}
P(T_{\rm C} > t) \simeq
1 - \exp \left[ - \langle U \rangle_t \right],
\label{eq:Gumbel2b}
\end{equation}

\noindent
where $\langle U \rangle_t$ is the mean number of yet-unvisited nodes at time $t$.

Rearranging terms in the exponent, it is found that
the distribution of cover times is a discrete Gumbel distribution,
known from extreme value theory,
which takes the form
\cite{Gumbel1935}

\begin{equation}
P(T_{\rm C} > t) \simeq
1 - \exp \left[ -   \exp \left( {- \frac{ t - \mu } {\beta } } \right) \right],
\label{eq:Gumbel3}
\end{equation}

\noindent
where

\begin{equation}
\mu =  2 + \frac{c-1}{c-2}  N \ln (N-2)
\end{equation}

\noindent
is called the location parameter and

\begin{equation}
\beta = \frac{c-1}{c-2} N 
\end{equation}

\noindent
is called the scale parameter.
The location parameter $\mu$ is equal to the mode of the Gumbel distribution.
The scale parameter $\beta$ is equal to the standard deviation up to a constant
factor of order $1$.

\noindent
The Gumbel distribution often emerges as the distribution of the 
maxima among sets of $n$ independent random variables drawn from the same distribution.
It is one of the three possible families of extreme value distributions
specified by the extreme value theory,
namely the Gumbel, Fr\'echet and Weibull families
\cite{Gumbel1935,Frechet1927,Fisher1928,Mises1936,Gnedenko1943}.
The Gumbel distribution appears in various problems that involve
structural and dynamical processes on random networks.
These include the distribution of diameters in an ensemble of subcritial ER networks
\cite{Luczak1998,Hartmann2018},
the distribution of the number of neighbors of a set of nodes
\cite{Rukhin2012,Zhukovskii2018,Rodionov2020},
the distribution of take-over times of infections
\cite{Ottino2017},
the distribution of extinction times of infections
\cite{Windridge2015}
and the distribution of flooding times
\cite{Hofstad2002,Gautreau2007}.

Using the point of view of the extreme-value theory, the distribution $P(T_{\rm C} \le t)$
can be considered as the distribution of the maximum among 
$N-2$ distributions of first passage times from the initial node $i$
to all the other nodes in the network (apart from the two nodes visited
in the first two time steps).
Therefore, under the assumption that the distributions of first passage times
for different nodes are independent, the distribution of cover times satisfies

\begin{equation}
P(T_{\rm C} > t) = 1 - [ P(T_{\rm FP} \le t) ]^{N-2}.
\label{eq:CoverFP}
\end{equation}

\noindent
Comparing Eqs. (\ref{eq:Gumbel1}) and (\ref{eq:CoverFP}), we conclude that
the distribution of first passage times is given by

\begin{equation}
P(T_{\rm FP} \le t) = 1 - e^{- \left( \frac{c-2}{c-1} \right) \frac{t-2}{N} }.
\end{equation}

\noindent
Indeed, the distribution of first passage times exhibits an exponential tail.
It thus meets the criterion for the emergence of the Gumbel distribution
in the Fisher-Tippet-Gnedenko theorem
\cite{Frechet1927,Fisher1928,Mises1936,Gnedenko1943}.
Note that the first passage times of adjacent target nodes may be correlated.
However, such correlations appear to have little effect on the distribution of
cover times.

The probability mass function of cover times 
can be obtained by taking the difference

\begin{equation}
P(T_{\rm C}=t) = 
P(T_{\rm C}>t-1) - P(T_{\rm C}>t).
\end{equation}

\noindent
Using Eq. (\ref{eq:PTCgt}) it is found that

\begin{equation}
P(T_{\rm C}=t) = 
P_{t}(S=N) - P_{t-1}(S=N).
\end{equation}
 
\noindent
Alternatively, the distribution of cover times can be 
expressed in the form

\begin{equation}
P(T_{\rm C}=t) = P_{t-1}(S=N-1) \Delta(N-1),
\label{eq:PTCtmf}
\end{equation}

\noindent
where, using Eq. (\ref{eq:Deltas}),

\begin{equation}
\Delta(N-1) = \frac{c-2}{(c-1)N}.
\end{equation}

\noindent
Eq. (\ref{eq:PTCtmf}) expresses the fact that in order for the
RW to enter the last unvisited node at time $t$ it has to visit
$N-1$ nodes up to time $t-1$ and step into a previously 
unvisited node at time $t$.

\begin{figure}
\centerline{
\includegraphics[width=14cm]{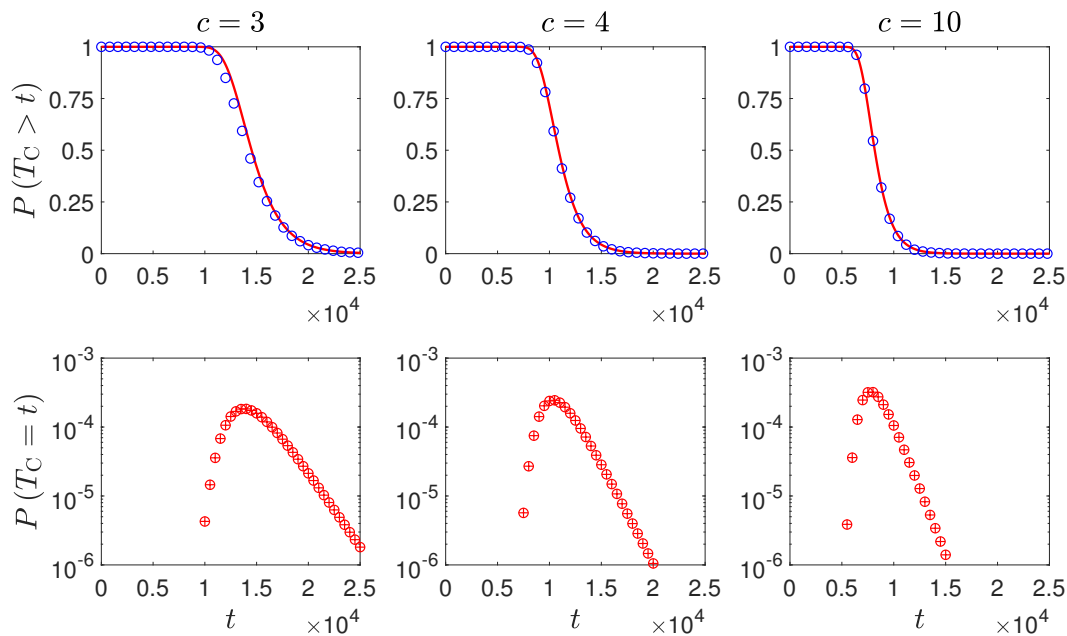}
}
\caption{
Top row: analytical results for the
tail distribution 
$P(T_{\rm C} > t)$  (solid lines)
of the cover times
of RWs on RRGs
of size $N=1000$ (top row)
and degree 
$c=3$ (left), $4$ (middle) and $10$ (right).
The analytical results, 
obtained from Eq. (\ref{eq:P_t(S=N)!2}) 
are in very good agreement with the results obtained
from computer simulations (circles).
Bottom row: the corresponding probability mass functions 
$P(T_{\rm C}=t)$,
obtained from Eq. (\ref{eq:PTCtmf}) (circles)
and from the Gumbel distribution 
[Eq. (\ref{eq:Gumbel3})] ($+$ symbols).
As the degree of the network increases,
the cover times become shorter and more concentrated around the mean.
}
\label{fig:7}
\end{figure}

In the top row of Fig. \ref{fig:7} 
we present the  tail distribution of cover times,
$P \left(T_{\rm C}>t\right)$ vs. $t$,
for an RRG of size $N=1000$ consisting of nodes of degree $c = 3$
(left), $4$ (middle) and $10$ (right). 
For $c=4$ and $c=10$
the analytical results (solid lines),
obtained from Eq.
(\ref{eq:P_t(S=N)!2}), 
are in very good agreement with the results obtained from computer simulations
(circles).
In the case of $c=3$ there is a slight discrepancy, where the analytical
results for the cover time are shifted to the right by a few hundred time steps 
compared to the simulation results.
This discrepancy is due to subtle correlations that emerge in low-degree RRGs,
which are most pronounced in the case of $c=3$.

In the bottom row of Fig. \ref{fig:7} we present analytical results for
the corresponding probability density functions 
$P \left(T_{\rm C}=t\right)$ vs. $t$. 
The results obtained from Eq. (\ref{eq:PTCtmf}) (circles)
are found to be in very good agreement with
the results obtained from the Gumbel distribution 
[Eq. (\ref{eq:Gumbel3})] ($+$ symbols).
It can be seen that as the mean degree is increased,
the cover times becomes shorter, and the distribution more centralized around
its mean. This is consistent with the general trend seen
in previous sections, namely the fact that
in sparser networks
the discovery of new nodes is generally slower compared to denser networks. As mentioned,
this can be explained by the fact that in low-degree networks there is a higher probability of backtracking to
previously visited domains of the network.
Note that for $t \le N-1$ the distribution of cover times satisfies $P(T_{\rm C}=t)=0$.
The corresponding Gumbel distribution is vanishingly small in this regime,
but not strictly zero.

\section{The mean cover time}

The mean cover time is given by

\begin{equation}
\langle T_{\rm C} \rangle =
\sum_{t=N}^{\infty} t P(T_{\rm C}=t).
\label{meanTc}
\end{equation}

\noindent
It can be expressed in the form

\begin{equation}
\langle T_{\rm C} \rangle =
\frac{d}{d \omega} J(\omega) \bigg|_{\omega=1},
\label{eq:TcJ}
\end{equation}

\noindent
where 

\begin{equation}
J(\omega) = 
\sum_{t=N}^{\infty} \omega^t P(T_{\rm C}=t) 
\label{eq:JPCT}
\end{equation}

\noindent
is the generating function of $P(T_{\rm C}=t)$.
In Appendix D we expand the generating function
$J(\omega)$ in powers of $\omega-1$.
Inserting the expansion of $J(\omega)$
from Eq. (\ref{eq:Jwexp}) into Eq. (\ref{eq:TcJ}), 
we obtain

\begin{equation}
\langle T_{\rm C} \rangle =
2 + \left( \frac{c-1}{c-2} \right) (N-2) H_{N-2},
\label{eq:LTCR}
\end{equation}

\noindent
where $H_{m}$ is the $m$th Harmonic number
\cite{Olver2010}.
In the limit of $m \gg 1$ the Harmonic numbers can be approximated by

\begin{equation}
H_m = \ln m + \gamma + \frac{1}{2m} + \mathcal{O} \left( \frac{1}{m^2} \right),
\label{eq:Harmonic_approx}
\end{equation}
 
\noindent
where $\gamma \simeq 0.577$ is the Euler-Mascheroni constant
\cite{Finch2003}.
Using this approximation, it is found that in the large network limit
the mean cover time can be expressed by

\begin{equation}
\langle T_{\rm C} \rangle =
\left( \frac{c-1}{c-2} \right) N \ln N
+ \left( \frac{c-1}{c-2}  \right) \gamma  N
+ \frac{c-5}{2(c-2)} + \mathcal{O} \left( \frac{1}{N} \right).
\label{eq:meanCovT}
\end{equation}

\noindent
The leading term on the right hand side of Eq. (\ref{eq:meanCovT})
coincides with the result
obtained in Ref.
\cite{Cooper2005}, using other methods.
Our solution also includes sub-leading terms which
can be significant even for  very large networks.

\begin{figure}
\centerline{
\includegraphics[width=8cm]{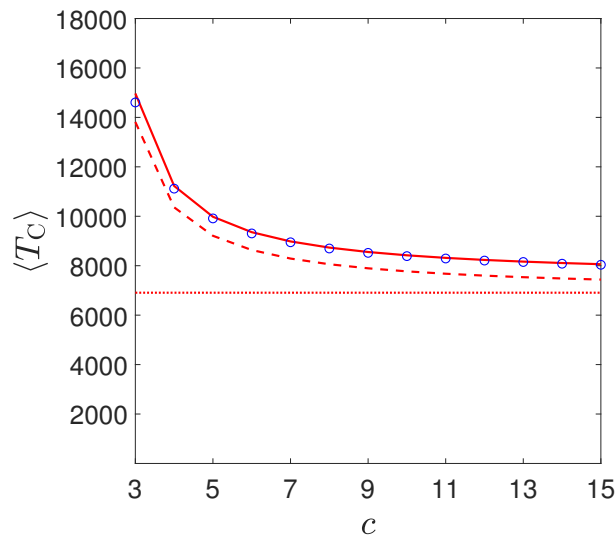}
}
\caption{
Analytical results for the
mean cover time 
$\langle T_{\rm C} \rangle$ (solid line)
for RRGs of size
$N=1000$
as a function of the degree $c$.
The analytical results,
obtained from Eq. (\ref{eq:LTCR}) 
are in very good agreement with the results
obtained from computer simulations (circles).
The leading order asymptotic approximation,  
obtained from Eq.
(\ref{eq:meanCovT}),
is also shown (dashed line).
The asymptotic value of the mean cover time
in the limit of high degree $c$, namely $N \ln N$, is also shown (dotted line).
}
\label{fig:8}
\end{figure}

In Fig. \ref{fig:8} we show the mean cover time vs node degree c,
for networks of size $N=1000$. The analytical result obtained from 
Eq. (\ref{eq:LTCR}) is shown in solid line, and the leading order asymptotic 
expression from Eq. (\ref{eq:meanCovT}) is shown in dashed line. 
For comparison, we also show the large $c$ limiting value $N \ln N$ in dotted line. 
The solid curve is
in good agreement with numerical simulations, and the dashed asymptotic 
limit also gives a good estimate, but with a noticeable discrepancy for networks of this size,
coming from the ${\mathcal O}(N)$ correction to the leading 
$\mathcal{O}(N \ln N)$
behavior.

\section{The variance of the distribution of cover times}

The second moment of the distribution of cover times can be expressed 
in terms of the generating function $J(\omega)$ in the form

\begin{equation}
\langle T_{\rm C}^2 \rangle = 
\frac{d^2}{d \omega^2} J(\omega) \bigg|_{\omega=1}
+
\frac{d}{d \omega} J(\omega) \bigg|_{\omega=1}.
\label{eq:T2cJ}
\end{equation}

\noindent
Inserting the expansion of $J(\omega)$
from Eq. (\ref{eq:Jwexp}) into Eq. (\ref{eq:T2cJ}), 
we obtain

\begin{equation}
\langle T_{\rm C}^2 \rangle =
4 + 3 \frac{c-1}{c-2} (N-2) H_{N-2} 
+ \left( \frac{c-1}{c-2} \right)^2  (N-2)^2   \left[ (H_{N-2})^2 + H_{N-2}^{(2)} \right],
\end{equation}

\noindent
where $H_{m}^{(2)}$
is the generalized Harmonic number of the second order
\cite{Olver2010}.
Combining the results for the first and second moments,
we obtain the variance of the distribution of cover times, 
which is given by

\begin{equation}
{\rm Var}(T_{\rm C}) =
\left( \frac{c-1}{c-2} \right)^2 (N-2)^2 H_{N-2}^{(2)}
- \left( \frac{c-1}{c-2} \right) (N-2) H_{N-2}.
\end{equation}

\noindent
In the limit of $N \gg 1$ the variance can be approximated by

\begin{eqnarray}
{\rm Var}(T_{\rm C}) =
\frac{\pi^2}{6}
\left( \frac{c-1}{c-2} \right)^2 N^2 
-
\left( \frac{c-1}{c-2} \right) N   \ln N
\nonumber \\
- \left( \frac{c-1}{c-2} \right) \left(   \gamma + \frac{c-1}{c-2} \right) N
- \frac{3(c-1)}{2(c-2)^2} + O \left(\frac{1}{N}\right).
\label{eq:VarTC}
\end{eqnarray}

\noindent
The leading term on the right hand side of Eq. (\ref{eq:VarTC})
scales like $N^2$, which means that the standard deviation of 
$P(T_{\rm C}=t)$ scales like $N$,
while the mean $\langle T_{\rm C} \rangle$ scales like $N \ln N$.

\begin{figure}
\centerline{
\includegraphics[width=8cm]{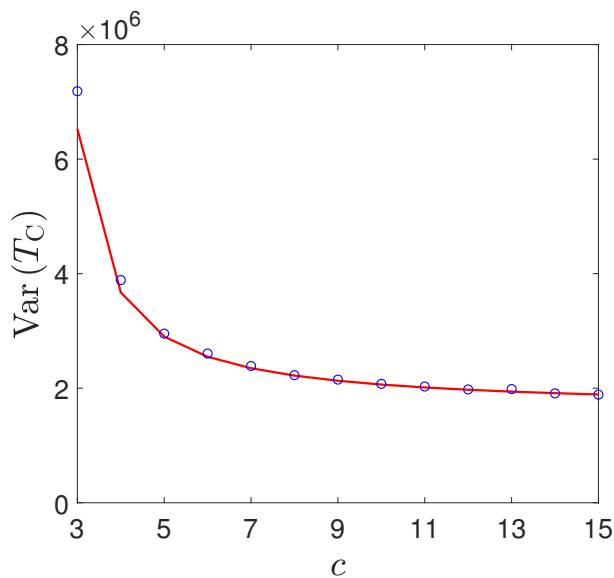}
}
\caption{
Analytical results for
the variance ${\rm Var}(T_{\rm C})$ 
of the distribution of cover times
as a function of the degree $c$ for an RW
on an RRG of size
$N=1000$ (solid line).
The analytical results,
obtained from Eq. (\ref{eq:VarTC}),
are in very good agreement with the results
obtained from computer simulations (circles),
except for the case of $c=3$ in which there
is a significant deviation.
}
\label{fig:9}
\end{figure}

In Fig. \ref{fig:9} we present
analytical results for
the variance ${\rm Var}(T_{\rm C})$ 
of the distribution of cover times
as a function of the degree $c$ for an RW
on an RRG of size
$N=1000$ (solid line).
The analytical results,
obtained from Eq. (\ref{eq:VarTC}),
are in very good agreement with the results
obtained from computer simulations (circles),
except for the case of $c=3$ in which there
is a significant deviation.

\section{The distribution of partial cover times}

One can generalize the concept of cover time to the $k$th order partial cover time,
which is the first time in which the RW visits $k$ distinct nodes
\cite{Nascimento2001,Coutinho1994,Chupeau2015}.
We denote the probability
that the RW will complete visiting $k$ distinct nodes at time $t$ 
by $P(T_{{\rm PC},k}=t)$.
This probability can be expressed in the form

\begin{equation}
P(T_{{\rm PC},k}=t) =
\left( \frac{c-2}{c-1} \right)
\left(1 - \frac{k-1}{N} \right)
P_{t-1}(S=k-1) ,
\label{eq:PTPCkt}
\end{equation}

\noindent
where $P_{t-1}(S=k-1)$ is the probability that at time $t-1$ the RW
has visited $k-1$ distinct nodes.
This probability is multiplied by the
probability that at time $t$ the RW will step into a new node,
which has not been visited before,
namely by $\Delta(k-1)$,
given by Eq. (\ref{eq:Deltas}).
Comparing Eq. (\ref{eq:PTPCkt}) with Eq. (\ref{eq:PTts}) in Appendix B, it is found that

\begin{equation}
P(T_{{\rm PC},k}=t) = P(T=t-1|k-1).
\label{eq:PTPCkt2}
\end{equation}

Below we present an identity that will be useful for the calculation of the
moments of the distribution of partial cover times.
Multiplying Eq. (\ref{eq:PTPCkt2}) by $t^r$ and summing over $t$,
we obtain

\begin{equation}
\langle T_{{\rm PC},k}^r \rangle = \mathbb{E}[(T+1)^r|S=k-1].
\label{eq:Tpckr}
\end{equation}

\noindent
Using Eq. (\ref{eq:ETs1}), we obtain the
mean of the $k$th order partial cover time, which is given by

\begin{equation}
\langle T_{{\rm PC},k} \rangle = 2 + \left( \frac{c-1}{c-2} \right) (N-2) (H_{N-2} - H_{N-k}).
\label{eq:TPCk}
\end{equation}

\noindent
Approximating the Harmonic numbers using Eq. (\ref{eq:Harmonic_approx}),
we obtain

\begin{equation}
\langle T_{{\rm PC},k} \rangle =
2 +
\frac{c-1}{c-2} N \ln \left( \frac{N-2}{N-k} \right)   + \mathcal{O} \left( \frac{1}{N} \right),
\label{eq:<TPC>}
\end{equation}

\noindent
where $2 \le k \le N-1$.
In the limit of $k \ll N$, Eq. (\ref{eq:<TPC>}) is reduced to

\begin{equation}
\langle T_{{\rm PC},k} \rangle \simeq
2 + \frac{c-1}{c-2} (k-2).
\label{eq:<TPC>b}
\end{equation}

\noindent
Note that Eq. (\ref{eq:<TPC>}) does not hold in the special case of $k=N$,
in which the partial cover time coincides with the cover time, namely
$\langle T_{{\rm PC},N} \rangle = \langle T_{\rm C} \rangle$.
Using Eq. (\ref{eq:Tpckr}) with $r=2$ and Eq. (\ref{eq:VTsb}), we obtain the
variance of the distribution of partial cover times, which takes the form

\begin{eqnarray}
{\rm Var}(T_{{\rm PC},k}) &=& 
\left( \frac{c-1}{c-2} \right)^2 (N-2)^2 \left[ H_{N-2}^{(2)} - H_{N-k}^{(2)} \right]
\nonumber \\
&-& \left( \frac{c-1}{c-2} \right) (N-2) (H_{N-2} - H_{N-k}).
\label{eq:VarTPCk}
\end{eqnarray}

\noindent
In the limit of $k \ll N$, Eq. (\ref{eq:VarTPCk}) is reduced to

\begin{equation}
{\rm Var}(T_{{\rm PC},k}) = 
\frac{c-1}{(c-2)^2} (k-2).
\label{eq:VarTPCkb}
\end{equation}

\section{The distribution of random cover times}

Another generalization of the concept of the cover time is referred to
as the random cover time. This is the first time at which the RW completes 
visiting a specific pre-selected set of $k$ randomly selected target nodes
\cite{Nascimento2001,Coutinho1994,Chupeau2015}.
The distribution of random cover times is denoted by $P(T_{{\rm RC},k}=t)$.
The tail distribution of random cover times can be expressed by 

\begin{equation}
P(T_{{\rm RC},k} > t) = 
1 - \sum_{s=k}^N
\frac{ \binom{N-k}{s-k} }{ \binom{N}{s} } 
P_t(S=s),
\label{eq:PTRCkt0}
\end{equation}

\noindent
where the ratio between the binomial coefficients
provides the probability that all the $k$ pre-selected nodes are included
in the $s$ distinct nodes visited by the RW up to time $t$.
The mean of the $k$th random cover time is given by
the tail-sum formula

\begin{equation}
\langle T_{{\rm RC},k} \rangle = 
\sum_{t=1}^{\infty}
P(T_{{\rm RC},k} > t-1).
\label{eq:<TRCk>}
\end{equation}

\noindent
Inserting $P(T_{{\rm RC},k} > t)$ from Eq. (\ref{eq:PTRCkt0})
into Eq. (\ref{eq:<TRCk>}), we obtain

\begin{equation}
\langle T_{{\rm RC},k} \rangle = 
\sum_{t=1}^{\infty}
\left[
1 -
\sum_{s=k}^N
\frac{ \binom{N-k}{s-k} }{ \binom{N}{s} } 
P_{t-1}(S=s)
\right].
\label{eq:Trck1}
\end{equation}

\noindent
It will be useful to introduce a generating function of the form

\begin{equation}
\rho_k(\omega)
=
\sum_{t=1}^{\infty} 
\omega^t P(T_{{\rm RC},k} \ge t).
\label{eq:rho}
\end{equation}

\noindent
Comparing Eqs. (\ref{eq:<TRCk>}) and (\ref{eq:rho}), it is found that the
mean $\langle T_{{\rm RC},k} \rangle$
can be expressed using this generating function, in the form

\begin{equation}
\langle T_{{\rm RC},k} \rangle = 
\lim_{\omega \rightarrow 1}
\rho_k(\omega).
\end{equation}

\noindent
Carrying out the summation in Eq. (\ref{eq:rho}),
we obtain

\begin{equation}
\rho_k(\omega) =
\frac{\omega}{1-\omega}
-
\omega \sum_{s=k}^{N}
\frac{ \binom{N-k}{s-k} }{ \binom{N}{s} }
L_s(\omega),
\label{eq:rho2}
\end{equation}

\noindent
where $L_s(\omega)$ is defined in Appendix A.
Being interested in the limit of $\omega \rightarrow 1$, 
we would like to expand the generating function $\rho_k(\omega)$
in powers of $\omega-1$.
Inserting $L_s(\omega)$ from Eqs. 
(\ref{eq:gammasw}) 
(for $s \le N-1$) and from (\ref{eq:gNom}) (for $s=N$),
we obtain

\begin{equation}
\rho_k(\omega) =
2
+ \left( \frac{c-1}{c-2} \right) (N-2)
\left[ H_{N-2}
- \sum_{s=k}^{N-1}
\frac{ \binom{N-k}{s-k} }{ \binom{N}{s} }
\left( \frac{1}{N-s} \right) \right]
+ \mathcal{O}(\omega-1).
\end{equation}

\noindent
Taking the limit of $\omega \rightarrow 1$
and carrying out the summation, we obtain

\begin{equation}
\langle T_{{\rm RC},k} \rangle = 
2 + (N-2) \left( \frac{c-1}{c-2} \right) 
\left[ H_{N-2} -  
(H_N - H_k)
\right].
\label{eq:Trck2}
\end{equation}

\noindent
After some algebraic simplifications,
we obtain a simple expression for the random cover time, 
which takes the form

\begin{equation}
\langle T_{{\rm RC},k} \rangle = 
\left( \frac{c-1}{c-2} \right) 
\left[ (N-2) H_{k} 
- \frac{2}{c-1} - \frac{1}{N-1}
\right].
\label{eq:Trck}
\end{equation}

\noindent
Note that the mean cover time $\langle T_{\rm C} \rangle$
can be recovered by inserting $k=N$ in Eq. (\ref{eq:Trck2}).
For values of $k$ which are not too small, one can approximate the
Harmonic number $H_k$ using Eq. (\ref{eq:Harmonic_approx}).
This leads to

\begin{equation}
\langle T_{{\rm RC},k} \rangle = 
\left( \frac{c-1}{c-2} \right) N \ln k + \left( \frac{c-1}{c-2} \right) \gamma N
+ \mathcal{O} \left( \frac{N}{k} \right).
\end{equation}

Below we show that for $t \gg N$ the distribution of random cover times follows a Gumbel distribution.
This will allow us to evaluate its variance ${\rm Var}(T_{{\rm RC},k})$.
Inserting $P_t(S=s)$ from Eq. (\ref{eq:PtSalt7}),
which is valid for $t \gg N$, into Eq. (\ref{eq:PTRCkt0}), 
we obtain

\begin{equation}
P(T_{{\rm RC},k} > t) = 
1 - \sum_{s=k}^N
\frac{ \binom{N-k}{s-k} }{ \binom{N}{s} } 
\binom{N-2}{s-2}
\left[ 1 - e^{ - \left( \frac{c-2}{c-1} \right) \frac{t-2}{N}  }   \right]^{s-2} 
\left[   e^{ - \left( \frac{c-2}{c-1} \right) \frac{t-2}{N}  }   \right]^{N-s}.
\label{eq:PTRCkt2z}
\end{equation}

\noindent
Eq. (\ref{eq:PTRCkt2z}) can be simplified to

\begin{eqnarray}
P(T_{{\rm RC},k} > t)  
&=& 
1 - \sum_{s=k}^N
\binom{N-k}{s-k} 
\frac{s(s-1)}{N(N-1)}
\times
\nonumber \\
& &
\ \ \ \ \ \ 
\left[ 1 - e^{ - \left( \frac{c-2}{c-1} \right) \frac{t-2}{N}  }  \right]^{s-2}
\left[   e^{ - \left( \frac{c-2}{c-1} \right) \frac{t-2}{N}  }   \right]^{N-s}.
\label{eq:PTRCkt3}
\end{eqnarray}

\noindent
Carrying out the summation, we obtain

\begin{eqnarray}
& & P(T_{{\rm RC},k} > t) = 
1 - \left[ 1 - e^{ - \left( \frac{c-2}{c-1} \right) \frac{t-2}{N}  }  \right]^{k} 
\times 
\nonumber \\
& &
\left\{ 1 + 2 \frac{k}{N}  
\left[ \frac{ e^{ - \left( \frac{c-2}{c-1} \right) \frac{t-2}{N}  } }
{1 - e^{ - \left( \frac{c-2}{c-1} \right) \frac{t-2}{N}  } } \right]
+
\frac{k(k-1)}{N(N-1)}
\left[ \frac{ e^{ - \left( \frac{c-2}{c-1} \right) \frac{t-2}{N}  } }
{1 - e^{ - \left( \frac{c-2}{c-1} \right) \frac{t-2}{N}  } } \right]^2
\right\}. 
\label{eq:PTRCkt4}
\end{eqnarray}

\noindent
In the limit of $t \gg N$ Eq. (\ref{eq:PTRCkt4}) can be approximated by

\begin{equation}
P(T_{{\rm RC},k} \le t)  =
1 - P(T_{{\rm RC},k}>t) =
\exp \left[ -  e^{ - \left( \frac{c-2}{c-1} \right) \frac{t-2- \frac{c-1}{c-2} N \ln k}{N}  }  \right],
\label{eq:PTRCkt}
\end{equation}

\noindent
which is a Gumbel distribution.
This distribution can be expressed by Eq. (\ref{eq:Gumbel3})
with 

\begin{equation}
\mu =  2 + \frac{c-1}{c-2}  N \ln k
\end{equation}

\noindent
and

\begin{equation}
\beta = \frac{c-1}{c-2} N.
\end{equation}

\noindent
The variance of the distribution of random cover times,
up to leading orders in $N$, 
is thus given by

\begin{equation}
{\rm Var}(T_{{\rm RC},k}) \simeq
\frac{\pi^2}{6}
\left( \frac{c-1}{c-2} \right)^2 N^2 
-
\left( \frac{c-1}{c-2} \right) N   \ln k.
\end{equation}

\noindent
The mean number of yet-unvisited nodes up to time $t$
in the random subgraph of $k$ nodes is given by

\begin{equation}
\langle U_k \rangle_t = \frac{k}{N} \langle U \rangle_t,
\label{eq:<U_k>_t}
\end{equation}

\noindent
where $\langle U \rangle_t$ is given by Eq. (\ref{eq:Utlate2}).
Thus, Eq. (\ref{eq:PTRCkt}) can be rewritten in the form

\begin{equation}
P(T_{{\rm RC},k} \le t)  = 
e^{- \langle U_k \rangle_t}.
\label{eq:PTRCkt2}
\end{equation}

\section{ The relation between $P(T_{{\rm PC},k}=t)$ and $P(T_{{\rm RC},k}=t)$ }

\begin{figure}
\centerline{
\includegraphics[width=8cm]{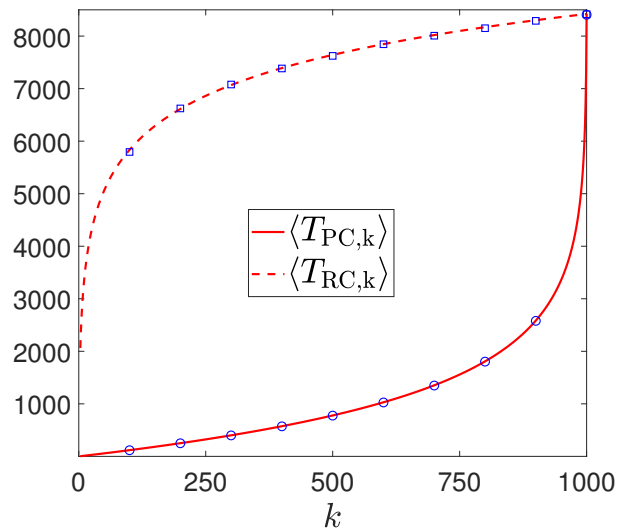}
}
\caption{
Analytical results for
the mean partial cover time
$\langle T_{ {\rm PC}, k} \rangle$ 
(solid line) 
and the mean random cover time $\langle T_{ {\rm RC},k} \rangle$ 
(dashed line), 
as a function of the number of nodes $k$,
for an RRG of size
$N=1000$ and degree $c=10$.
The analytical results,
obtained from Eq.
(\ref{eq:TPCk}) and Eq. (\ref{eq:Trck}) respectively,
are in very good agreement with the results obtained from computer 
simulations 
(circles and squares, respectively).
Note that the two curves exhibit a reflection symmetry between them,
as discussed in the main text.
}
\label{fig:10}
\end{figure}

In Fig. \ref{fig:10} 
we present the mean partial cover time 
$\langle T_{{\rm PC},k}   \rangle$, as a function of $k$, in a network 
of size $N=1000$ and node degree $c=10$. The analytical results (solid line) 
given by Eq. (\ref{eq:TPCk}), 
are found to be in very good agreement with the results
obtained from simulations (symbols).
It can be seen that the mean partial cover time grows very rapidly only when
$k$ approaches $N$. 
We also present the mean random cover time  
$\langle T_{{\rm RC},k} \rangle$ 
vs. $k$.
The analytical result (solid line) 
given by Eq. (\ref{eq:Trck}),  is found to 
be in very good agreement with the results obtained from computer simulations (symbols). 
It is found that for any value of $1 \le k < N$ the random cover time
$\langle T_{{\rm RC},k} \rangle$
is much larger than the corresponding partial cover time
$\langle T_{{\rm PC},k} \rangle$.
This reflects the fact that it takes longer to visit $k$ pre-selected nodes than to
visit a set of $k$ unspecified nodes.

Fig. \ref{fig:10} reveals a surprising reflection symmetry between
$\langle T_{{\rm RC},k} \rangle$ and $\langle T_{{\rm PC},k} \rangle$.
Taking each one of these functions and applying the inversions
$k \rightarrow N - k$ and 
$\langle T \rangle \rightarrow \langle T_{\rm C} \rangle - \langle T \rangle$
one obtains the other functions up to a very small shift.
The reflection symmetry can be expressed by

\begin{equation}
\langle T_{{\rm RC},N-k} \rangle + \langle T_{{\rm PC},k} \rangle =
\langle T_{\rm C} \rangle  
+ \frac{2}{c-2} + \left( \frac{c-1}{c-2} \right) \frac{1}{N-1}.
\end{equation}

\noindent
Note that the two correction terms on the right hand side 
do not depend on $k$. Moreover, the first correction term is
of the order of $1/c$ while the second term is of the order of $1/N$.
Both are negligible compared to $\langle T_C \rangle$, which scales like
$N \ln N$.
Thus, for sufficiently large networks

\begin{equation}
\langle T_{{\rm RC},N-k} \rangle + \langle T_{{\rm PC},k} \rangle =
\langle T_{\rm C} \rangle. 
\end{equation}

\noindent
This symmetry reflects the following property: 
starting the RW from some
random node $i$, 
the mean number of time steps it will take to visit $k$ distinct nodes 
(which were not specified beforehand)
is given
by $\langle T_{{\rm PC},k} \rangle$.
At this stage there are $N-k$ remaining nodes to cover in order to 
complete the cover time. However, the remaining $N-k$ nodes are
specific ones, because these are the nodes that have not been visited up to 
that time. Therefore, the time that will take the RW to cover the remaining 
$N-k$ nodes follows the distribution of random cover times of $N-k$ nodes,
whose mean is $\langle T_{{\rm RC},N-k} \rangle$.

To put the three types of cover times on a common footing, we summarize
their scaling behavior:
the mean of the distribution of cover times scales like
$\langle T_{\rm C} \rangle \sim N \ln N$,
the mean of the distribution of partial cover times scales like
$\langle T_{{\rm PC},k} \rangle \sim N \ln \left( \frac{N}{N-k} \right)$
and the mean of the distribution of random cover times scales like
$\langle T_{{\rm RC},k} \rangle \sim N \ln k$.

\section{Discussion}

A characteristic property of the cover time problem is that at early times the RW
is highly efficient in covering new nodes. This efficiency is gradually reduced
as the fraction of nodes that have already been visited increases,
until at late times it takes a large number of steps to reach each one of the
few yet-unvisited nodes that remain.
Such situations are often described by the 80/20 law (or Pareto principle),
which states that in certain systems roughly 80 percent of the outcome is a result of only 20
percent of the effort, while the remaining 20 percent or so of the outcome consumes 80 percent
of the effort
\cite{Pareto1898,Amoroso1938}.
In light of this observation, it is interesting to find the value of $0 < f < 1$
for which $f \langle T_{\rm C} \rangle$ time steps of an RW would cover,
on average, $(1-f)N$ nodes of the RRG.
The fraction $f$ can be calculated by solving the equation

\begin{equation}
\langle T_{ {\rm PC},(1-f)N } \rangle = f \langle T_{\rm C} \rangle.
\label{eq:Pareto}
\end{equation}

\noindent
Taking the large $N$ limit, we insert
the leading term for
$\langle T_{ {\rm PC},(1-f)N } \rangle$ from Eq. (\ref{eq:<TPC>})
and for $\langle T_{\rm C} \rangle$ from Eq. (\ref{eq:meanCovT}) into 
Eq. (\ref{eq:Pareto}).
We obtain

\begin{equation}
(1-f) \ln N = \ln (f N) + \gamma f.
\label{eq:fNfN}
\end{equation}

\noindent
Eq. (\ref{eq:fNfN}) can be written in the form

\begin{equation}
f N^{f} = e^{-\gamma f},
\end{equation}

\noindent
which implies that

\begin{equation}
f = \frac{ W(\gamma + \ln N) }{\gamma + \ln N },
\end{equation}

\noindent
where $W(x)$ is the Lambert W function
\cite{Olver2010}.
The fraction $f$ is a monotonically decreasing function of $N$.
For example, in case that $N=1000$ we obtain $f \simeq 0.21$.
This is consistent with the $80/20$ law.

The declining efficiency of the process of covering the network as
the time evolves is reminiscent of the economic law of diminishing returns
\cite{Cannan1892,Lloyd1969}.
Consider the production process of a commodity, in which
a single input component is increased 
while all the other input components are held fixed.
The law states that at some point the 
resulting increase in the output per unit increase in the input
will become progressively smaller or diminishing.
In this analogy the time steps of the RW are considered as the input resource
and the number of distinct nodes visited by the RW is the output or product.
As the time evolves the number of distinct nodes visited by
the RW per time step diminishes.
The practice of working on a task past the point of diminishing returns is 
often referred to as gold plating
\cite{McConnell2014}.
While the economic literature focuses on the negative side of gold plating, 
there are often great advantages and importance in bringing things to perfection or completion.

It is interesting to compare the results obtained in this paper 
for the cover times of RWs on RRGs 
with the corresponding results for RWs on regular lattices with the 
same coordination numbers.
For example, the coordination number of a hypercubic lattice in $d$
dimensions is $2d$. Thus, in terms of the connectivity 
the $d$-dimensional hypercubic lattice is
analogous to an RRG of degree $c=2d$.
In the case of an RW on a one-dimensional lattice of $N$ sites with periodic boundaries,
it was found that the mean cover time is given by
$\langle T_{\rm C} \rangle = N(N-1)/2$
\cite{Nemirovsky1990,Yokoi1990}.
For an RW on a two-dimensional square lattice consisting of
$N=L^2$ sites with periodic boundaries (forming a torus),
it was found that
$\langle T_{\rm C} \rangle \propto N (\ln N)^2$
\cite{Nemirovsky1990,Dembo2004,Grassberger2017}.
For dimensions $d \ge 3$ it was found that the mean cover time
of an RW on a cubic lattice consisting of $N=L^d$ sites with 
periodic boundaries is given by 
$\langle T_{\rm C} \rangle = A_d N \ln N$,
where the coefficient $A_d$ depends on the dimension $d$
\cite{Nemirovsky1990}.
Thus, for $d \ge 3$ the leading term in the expression for
$\langle T_{\rm C} \rangle$
on regular lattices has the same functional form as 
in the case of RRGs.
However, the values of the coefficient $A_d$, $d=3,4,\dots$, 
for regular lattices are known only approximately from computer simulations.
For example, it was found that
$A_3 \simeq 1.63$ and $A_4 \simeq 1.23$.
These values are larger than the coefficient obtained for
the corresponding RRG with $c=2d$, which is given by
$(c-1)/(c-2)$.
This implies that the cover time of an RW on a regular lattice
is larger than the cover time on an RRG with the same coordination 
number. This is sensible in light of the fact that the RRG is a small world
network on which it is less likely that the RW will remain for a long time
in the same neighborhood and revisit the same nodes again and again.
Interestingly, beyond any differences in the prefactor of the mean cover time, 
the limit distribution of cover times on RRGs and lattices with $d \ge 3$ is Gumbel in both cases
\cite{Belius2013}.
Based on the experience gained in the study of other problems on RRGs
\cite{Tishby2021b,Pathria2011,Plischke2006}
we expect
that the results for $\langle T_C \rangle$ provide the asymptotic large $d$ behavior
on regular lattices.
More precisely,
we conjecture that for hypercubic lattices of high dimension $d$ the coefficient $A_d$ is given by
$A_d \simeq (2d-1)/(2d-2)$.

Another type of random walk model is the non-backtracking random walk (NBW)
\cite{Alon2007,Tishby2017b}.
At each time step the NBW hops from its present node to one of its neighbors,
except for the node it visited in the previous time step.
It is thus similar to the RW, except for the backtracking step which is eliminated.
As a result, all the subsequent retroceding steps are also eliminated.
The paths of NBWs have been studied on regular lattices and random graphs 
\cite{Alon2007}.
It was shown that they explore the network more efficiently than RWs. 
The elimination of the backtracking step implies that on RRGs in
the $N \rightarrow \infty$ limit the probability that an NBW will step into
a yet unvisited node is $\Delta=1$.
Therefore, in the case of NBWs Eq. (\ref{eq:Deltas}) is replaced by

\begin{equation}
\Delta(s) =  1 - \frac{s}{N}.
\label{eq:DeltasNBW}
\end{equation}

\noindent
As a result, the distribution of cover times of NBWs on RRGs is given by
Eq. (\ref{eq:P_t(S=N)!2}) where $(c-2)/(c-1)$ is replaced by $1$.
In the long time limit $t \gg N$ it can be  approximated by the Gumbel
distribution, which is given by Eq. (\ref{eq:Gumbel3}) with $\mu = N \ln N$ and $\beta=N$.

Another interesting direction in the context of exploration of networks
using RWs is that of edge coverage
\cite{Costa2007}.
A typical problem is to determine the number of distinct edges visited by 
an RW up to time $t$.
The time it takes an RW on an RRG to visit every single edge in the
network is called the edge cover time. The distribution $P(T^{e}_{\rm C}=t)$ of edge cover times
of RWs on RRGs can also be calculated using the approach developed in this paper.
The number of edges in an RRG that consists of $N$ nodes of degree $c$ is
$N_e=Nc/2$.
The probability that an RW that has already visited $s_e$ distinct edges will step into
a yet-unvisited edge in the next time step is given by

\begin{equation}
\Delta_e(s_e) = \frac{c-2}{c-1} \left( 1 - \frac{s_e}{N_e} \right).
\end{equation}

\noindent
Thus, the distribution $P_t(S_e=s_e)$ of distinct edges visited by an RW up to time $t$ can be expressed by
Eq. (\ref{eq:P_t(S=s)!s}) or by Eq. (\ref{eq:PtSalt}) 
where $N$ is replaced by $N_e$.
Similarly, the tail distribution of edge cover times $P(T_{\rm C}^e>t)$
is given by Eq. (\ref{eq:P_t(S=N)!2}), where $N$ is replaced by $N_e$.

The cover time problem was studied for a broad range of 
random search processes 
\cite{Chupeau2015}. 
These search processes correspond to 
generalized random walk models such as L\'evy walks, intermittent
RWs and persistent RWs. It was shown that in all these systems
the distribution of cover times follows the Gumbel distribution.
It was thus concluded that the Gumbel distribution is a universal
distribution of cover times. 
Interestingly, it was recently shown that by accelerating
the search process one can modify
the distribution of cover times from the Gumbel distribution
to narrower distributions such as the Gaussian distribution
\cite{Cocconi2020}.

An interesting strategy for accelerating the covering of a network is by
using $k$ independent RWs
\cite{Alon2008,Alon2011,Kumar2019}.
In particular, it was shown rigorously that on random networks, as long as the number of RWs
is not too large, namely 
$k < (\ln N)^{1-\epsilon}$
(where $\epsilon$ is a small number),
the acceleration is by a factor of at least $k$.
In simple terms, the meaning of this is that a low concentration
of RWs overlap very mildly.

\section{Summary}

We presented analytical results for the distribution of
cover times of RWs on RRGs consisting of $N$ nodes of 
degree $c \ge 3$. 
To this end, we 
derived a master equation for the distribution
$P_t(S=s)$ of the number of distinct nodes $s$ visited by an RW up to time $t$. 
Using a generating function formalism, we 
solved the master equation and
obtained a closed-form analytical expression
for $P_t(S=s)$. 
Applying this result to the special case of $s=N$, we obtained
the cumulative distribution of cover times 
$P( T_{\rm C} \le t)=P_t(S=N)$ 
and calculated its mean and variance.
Taking the large network limit, we showed that
the distribution of cover times
follows a Gumbel distribution.
We also studied two interesting generalizations of the cover time:
the partial cover time 
$T_{{\rm PC},k}$, which is the
time it takes an RW to visit $k$ distinct nodes and the random cover time 
$T_{{\rm RC},k}$,
which is the time it takes an RW to cover a set of $k$ random pre-selected nodes.
The analytical results were compared to the results 
obtained from computer simulations and found to be in 
very good agreement.

We thank C. Cooper, S.N. Dorogovtsev, A.M. Frieze, N. Masuda and P. Sollich
for useful discussions.
This work was supported by the Israel Science Foundation grant no. 
1682/18.

\appendix

\section{Solution of the master equation for $P_t(S=s)$}

The discrete master equation [Eq. (\ref{eq:pnt0b})] consists of $N-1$ coupled difference equations
for the discrete time derivative of $P_t(S=s)$, $s=2,\dots,N$ at $t \ge 2$.
The initial condition is given by
$P_{2}(S=s)=\delta_{s,2}$.  
Note also that in the first time step
$P_{1}(S=s)=\delta_{s,1}$,
while at $t \ge 2$ the probability
$P_{t}(S=1)=0$.
The master equation can also be written in the form

\begin{eqnarray}
D_t P_t(S=s)  &=&
\left( \frac{c-2}{c-1} \right)
\left\{
\bigg[ P_t(S=s-1) - P_t(S=s) \bigg]
\right.
\nonumber \\
&-&  
\left.
\bigg[ \frac{s-1}{N}    P_{t}(S=s-1)
-
\frac{s}{N} P_{t}(S=s) 
\bigg]
\right\}.
\label{eq:re}
\end{eqnarray}

\noindent
The generating function of $P_t(S=s)$,
is given by

\begin{equation}
G_t(x) = \sum_{s=2}^N x^s P_t(S=s).
\label{eq:Gtxdef}
\end{equation}

\noindent
The generating function $G_t(x)$ represents a discrete Laplace transform of $P_t(S=s)$
with respect to $s$.
Multiplying both sides of Eq. (\ref{eq:re}) by $x^s$ and
summing up over $s$ we obtain

\begin{equation}
D_t G_t(x) =
\left( \frac{ c-2 }{ c-1 } \right) (x-1)  
\left[ G_t(x) - \frac{x}{N} \frac{ \partial }{ \partial x } G_t(x) \right].
\label{eq:Gpde}
\end{equation}

\noindent
The initial condition (at $t=2$), expressed in terms of the generating function $G_t(x)$,
takes the form

\begin{equation}
G_{2}(x) = \sum_{s=2}^N x^s P_2(S=s) = x^2.
\end{equation}

\noindent
We now define a second generating function, of the form

\begin{equation}
GL(x,\omega)
= \sum_{t=2}^{\infty}
\omega^t G_t(x).
\label{eq:Gamma_xo}
\end{equation}

\noindent
The generating function
$GL(x,\omega)$ 
represents a discrete double Laplace transform of $P_t(S=s)$
with respect to both $s$ and $t$.
Multiplying Eq. (\ref{eq:Gpde})
by $\omega^t$ and summing up over $t \ge 2$,
we obtain

\begin{equation}
(1-\omega) GL(x,\omega) - \omega^2 x^2 =
\left( \frac{c-1}{c-2} \right) (x-1) \omega
\left[ GL(x,\omega) - \frac{x}{N} \frac{\partial}{\partial x} GL(x,\omega) \right].
\label{eq:1oGxo}
\end{equation}

\noindent
Inserting $x=1$ in Eq. (\ref{eq:1oGxo}), the right hand side vanishes.
Solving the resulting equation, we obtain

\begin{equation}
GL(1,\omega)=\frac{\omega^2}{1-\omega}.
\label{eq:Gamma1}
\end{equation}

\noindent
Inserting $x=1$ in Eq. (\ref{eq:Gamma_xo}) and comparing the
result with Eq. (\ref{eq:Gamma1}), one can verify that $G_t(1)=1$,
as expected.
Using Eq. (\ref{eq:Gtxdef}),
this result establishes the normalization of $P_t(S=s)$.
In order to solve Eq. (\ref{eq:1oGxo}), we consider a series expansion
in powers of $x-1$, which is given by 

\begin{equation}
GL(x,\omega) = 
\frac{ \omega^2 x^2 }{ 1 - \omega }
\sum_{m=0}^{\infty}
a_m(\omega) (x-1)^m.
\label{eq:Gammasol}
\end{equation}

\noindent
Inserting $x=1$ in Eq. (\ref{eq:Gammasol}) and comparing the
result to Eq. (\ref{eq:Gamma1}), it is found that $a_0(\omega)=1$.
Inserting $GL(x,\omega)$ from Eq. (\ref{eq:Gammasol}) into Eq. (\ref{eq:1oGxo}),
we obtain recursion equations for the coefficients $a_m(\omega)$,
which take the form

\begin{equation}
a_m(\omega) = \left[ \frac{N-1-m}{m + \frac{1}{A(\omega)}} \right] a_{m-1}(\omega),
\label{eq:a_m}
\end{equation}

\noindent
where

\begin{equation}
A(\omega) = \left[ \left( \frac{c-2}{c-1} \right) \frac{1}{N} \right] \left( \frac{\omega}{1-\omega} \right) .
\label{eq:A}
\end{equation}

\noindent
From Eq. (\ref{eq:a_m}) it is clear that the recursion equations
terminate at $m=N-1$, and therefore $a_{N-1}(\omega)=0$.
Iterating the recursion equations (\ref{eq:a_m}),
starting from $a_0(\omega)=1$,
we obtain an explicit solution for the coefficients, which is given by

\begin{equation}
a_m(\omega) = \frac{ (-1)^m (2-N)_m }{\left( 1 + \frac{1}{A(\omega)} \right)_m},
\label{eq:a_m2}
\end{equation}

\noindent
where 

\begin{equation}
(b)_n=b(b+1) \dots (b+n-1) 
\label{eq:Poch}
\end{equation}

\noindent
is the (rising) Pochhammer symbol
\cite{Olver2010}.
Expressing the negative Pochhammer symbol
$(2-N)_m$
in Eq. (\ref{eq:a_m2})
in terms of a ratio of two factorials and
rearranging terms, one can express the coefficients $a_m(\omega)$
in the form

\begin{equation}
a_m(\omega) =  
\frac{(N-2)!}{(N-m-2)!} 
\prod_{\ell=1}^m
\frac{ 1 }{    \ell + \frac{1}{A(\omega)} }.
\label{eq:a_m2b}
\end{equation}

\noindent
Inserting the coefficients $a_m(\omega)$ from Eq. (\ref{eq:a_m2})
into Eq. (\ref{eq:Gammasol}), we obtain

\begin{equation}
GL(x,\omega) = \frac{ \omega^2 x^2 }{1 - \omega}  
 \ _2 F_1 \left[ \left.
\begin{array}{c}
1,2-N \\
1 + \frac{1}{A(\omega)}  
\end{array}
\right| 1-x
\right],
\label{eq:GammaPoch}
\end{equation}

\noindent
where the function
$_2F_1[ \ ]$
is the hypergeometric function 
\cite{Olver2010},
which is given by

\begin{equation}
_2F_1 \left[ \left.
\begin{array}{c}
a, b \\
d
\end{array}
\right| z 
\right] =
\sum_{n=0}^{\infty} 
\frac{ (a)_n (b)_n }{ (d)_n } \frac{ z^n }{ n! }.
\label{eq:2F1}
\end{equation}

\noindent
Expanding 
the right hand side of Eq. (\ref{eq:GammaPoch})
in a Taylor series around $x=0$,
we obtain

\begin{equation}
GL(x,\omega) = 
\frac{ \omega^2 }{ 1-\omega } 
\sum_{s=2}^{N} x^s 
\frac{ \Gamma \left( 1 + \frac{1}{A(\omega)}   \right) }{ \Gamma \left(  s - 1 + \frac{1}{A(\omega)} \right) }
\frac{ (N-2)! }{ (N-s)! } 
\ _2F_1 \left[ \left.
\begin{array}{c}
s-1,s-N \\
s-1+\frac{1}{A(\omega)}
\end{array}
\right| 1
\right],
\label{eq:GammaTaylor}
\end{equation}

\noindent
where $\Gamma(x)$ is the Gamma function
\cite{Olver2010}.

In order to obtain the distribution $P_t(S=s)$ we perform below
two inverse Laplace transforms, from $x$ to $s$ and from $\omega$ to $t$.
We first carry out the Laplace transform from $GL(x,\omega)$ to 
the generating function

\begin{equation}
L_s(\omega) = \sum_{t=2}^{\infty}
 \omega^t  P_t(S=s).
\label{eq:gsomega}
\end{equation}

\noindent
This generating function is related to $GL(x,\omega)$ via
the relation

\begin{equation}
GL(x,\omega) = \sum_{s=2}^{\infty}
x^s L_s(\omega).
\label{eq:Gx_gs}
\end{equation}

\begin{figure}
\centerline{
\includegraphics[width=6cm]{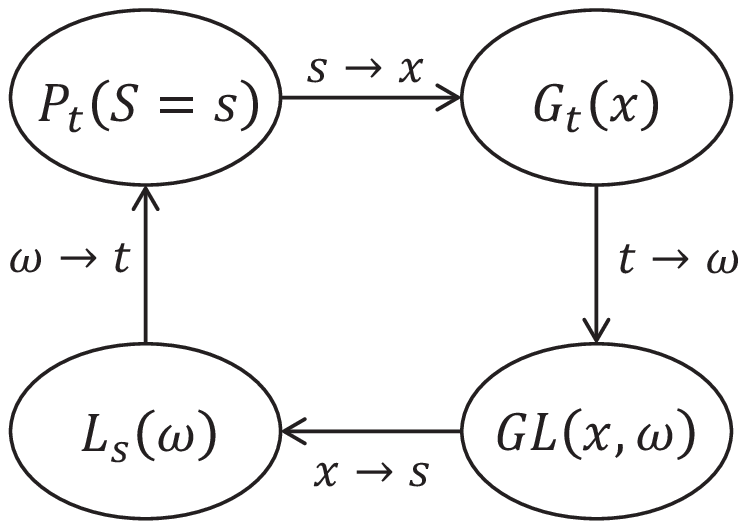}
}
\caption{
Illustration of the sequence of two discrete Laplace transforms followed by two inverse Laplace transforms,
that are used in the solution of the discrete master equation, 
providing a closed form expression for $P_t(S=s)$.
}
\label{fig:11}
\end{figure}

In Fig. \ref{fig:11} we illustrate the sequence of discrete Laplace transforms 
that is used in this Appendix for the calculation of $P_t(S=s)$. 
The sequence takes the form

\begin{equation}
P_t(S=s) \rightarrow G_t(x) \rightarrow GL(x,\omega) \rightarrow L_s(\omega) \rightarrow  P_t(S=s).
\end{equation}

\noindent
It consists of four functions of two variables each and four Laplace or inverse Laplace transformations between them.
The conjugate variable to the discrete time $t$ is the continuous variable $\omega$,
while the conjugate variable to the discrete variable $s$ is the continuous variable $x$.
It illustrates the sequence of steps pursued in the solution of the master equation.
The formal structure of these functions and the relations between them
are reminiscent of the four thermodynamic potentials  
[namely $U(S,V)$, $F(T,V)$, $G(T,P)$ and $H(S,P)$], 
related to each other by Legendre transforms
\cite{Pathria2011,Plischke2006}.

Comparing the coefficients of $x^s$ in Eqs. (\ref{eq:GammaTaylor}) and
(\ref{eq:Gx_gs}), we obtain

\begin{equation}
L_s(\omega) = 
\frac{ (N-2)! }{ (N-s)! }
\frac{ \Gamma \left( 1 + \frac{1}{A(\omega)} \right) }{ \Gamma \left( s - 1 + \frac{1}{A(\omega)} \right) }
\left( \frac{ \omega^2 }{ 1-\omega } \right)
\ _2F_1 \left[ \left.
\begin{array}{c}
s-1,s-N \\
s-1+\frac{1}{A(\omega)}
\end{array}
\right| 1
\right].
\label{eq:gammaTaylor}
\end{equation}

\noindent
Since one of the parameters in $_2F_1[ \ ]$
on the right hand side of 
Eq. (\ref{eq:gammaTaylor})
is a negative integer, the
hypergeometric function 
can be calculated using the identity

\begin{equation}
_2F_1 \left[ \left.
\begin{array}{c}
a,-m \\
d
\end{array}
\right| 1
\right]
=
\frac{ (d-a)_m }{(d)_m},
\label{eq:hypergem}
\end{equation}

\noindent
where $(d)_m$ is the Pochhammer symbol, given by Eq. (\ref{eq:Poch}), and $m$ is a positive integer.
This result is similar to the Chu-Vandermonde Identity  
(equation 15.4.24 in Ref. \cite{Olver2010}).
In the following we will need a few more results concerning this hypergeometric function.
First, in the limit of $a \rightarrow d$, we obtain

\begin{equation}
\,  
_2F_1 \left[ \left.
\begin{array}{c}
a,-m \\
d
\end{array}
\right| 1
\right] 
\xrightarrow[a \rightarrow d]{}  \frac{(0)_m}{(d)_m} = \delta_{m,0}.
\label{eq:2F1lim}
\end{equation}

\noindent
Also, the derivative of the hypergeometric function with respect to $d$ satisfies

\begin{equation}
\frac{\partial}{\partial d}
\,  
_2F_1 \left[ \left.
\begin{array}{c}
a,-m \\
d
\end{array}
\right| 1
\right] 
\xrightarrow[a \rightarrow d]{}
\frac{ \Gamma(d) \Gamma(m) }{\Gamma(d+m)},
\end{equation}

\noindent
while the second derivative satisfies

\begin{equation}
\frac{\partial^2}{\partial d^2}
\,  
_2F_1 \left[ \left.
\begin{array}{c}
a,-m \\
d
\end{array}
\right| 1
\right] 
\xrightarrow[a \rightarrow d]{} 
2 \frac{ \Gamma(d) \Gamma(m) }{\Gamma(d+m)}
\left( H_{m-1} + H_{d-1} - H_{m+d-1} \right),
\end{equation}

\noindent
where $H_m$ is the $m$th harmonic number
\cite{Olver2010}.
The third derivative satisfies

\begin{eqnarray}
\frac{\partial^3}{\partial d^3}
\,  
_2F_1 \left[ \left.
\begin{array}{c}
a,-m \\
d
\end{array}
\right| 1
\right] 
& \xrightarrow[a \rightarrow d]{} &
3 \frac{ \Gamma(d) \Gamma(m) }{\Gamma(d+m)}
\left[
\left( H_{m-1} + H_{d-1} - H_{m+d-1} \right)^2  \right.
\nonumber \\
& & -
\left. H_{m-1}^{(2)} - H_{d-1}^{(2)} + H_{m+d-1}^{(2)} \right],
\label{eq:d3F}
\end{eqnarray}

\noindent
where $H_{m}^{(2)}$ is the $m$th generalized harmonic number of the
second order
\cite{Olver2010}, 
which is also expressible in terms of the Riemann $\zeta$ function
\cite{Olver2010}.
Applying these results to Eq. (\ref{eq:gammaTaylor}), we obtain

\begin{equation}
L_s(\omega) =
\omega \left( \frac{c-1}{c-2} \right) 
\left(  1-\frac{s}{N}  \right)^{-1}
\frac{ \Gamma(N-1) }{ \Gamma(N-s) }
\frac{ \Gamma \left[ N-s+\frac{1}{A(\omega)} \right] }{ \Gamma \left[ N - 1 + \frac{1}{A(\omega)} \right]   }.
\label{eq:gsom2}
\end{equation}

\noindent
Expanding the expression on the right hand side of Eq. (\ref{eq:gsom2})
around $\omega=1$ for $s \le N-1$,
we obtain

\begin{eqnarray}
L_s(\omega) &=&
\left( \frac{c-1}{c-2} \right) \left(  1 - \frac{s}{N}  \right)^{-1}
\nonumber \\
&+&
\left( \frac{c-1}{c-2} \right) \left(  1 - \frac{s}{N} \right)^{-1}
\left[ 1 + \left( \frac{c-1}{c-2} \right) N (H_{N-2}-H_{N-s-1}) \right]
(\omega-1)  
\nonumber \\
&+& \mathcal{O} \left[ (\omega-1)^2 \right].
\label{eq:gammasw}
\end{eqnarray}

\noindent
Eq. (\ref{eq:gammasw}) provides the generating function $L_s(\omega)$
as a series in powers of $\omega-1$,
which will be useful in Appendix B.
In order to extract the distribution $P_t(S=s)$ we need to express
$L_s(\omega)$
as a series  
in powers of $\omega$. 

Consider the identity  
\cite{Boya2018}
 
\begin{equation}
(-1)^m
\prod_{\ell=1}^m
\frac{ 1 }{    \ell + \frac{1}{A(\omega)} }
=
\sum_{v=m}^{\infty}
\Big\{
\begin{array}{l}
v  \\
m       
\end{array}
\Big\}
[-A(\omega)]^{v},
\label{eq:Boya}
\end{equation}
 
\noindent
where 
$\Big\{
\begin{array}{l}
v  \\
m       
\end{array}
\Big\}$
is the Stirling number of the second kind.
Inserting Eq. (\ref{eq:Boya})
into Eq. (\ref{eq:a_m2b}),
we obtain

\begin{equation}
a_m(\omega) =  
(-1)^m
\frac{(N-2)!}{(N-m-2)!} 
%\prod_{\ell=1}^m
%\frac{ 1 }{    \ell + \frac{1}{A(\omega)} }.
\sum_{v=m}^{\infty}
\Big\{
\begin{array}{l}
v  \\
m       
\end{array}
\Big\}
[-A(\omega)]^{v}.
\label{eq:a_m2c}
\end{equation}

\noindent
Plugging Eq. (\ref{eq:a_m2c}) into Eq. (\ref{eq:Gammasol})
and expanding in powers of $x$, we obtain

\begin{equation}
L_s(\omega) =
\frac{ \omega^2 }{1-\omega}
\sum_{v=s-2}^{\infty}
(-1)^{v-s}
[A(\omega)]^{v} 
\sum_{m=s-2}^{\min\{v,N-2\}} 
m!
\Big\{
\begin{array}{l}
v  \\
m       
\end{array}
\Big\}
\binom{N-2}{m}
\binom{m}{s-2}.
\label{eq:gamma_s_omega}
\end{equation}

\noindent
In order to express $[A(\omega)]^{v}$ in terms of powers of $\omega$,
we use the Binomial identity
(equation 26.3.4 in Ref. \cite{Olver2010})

\begin{equation}
\frac{ 1 }{ (1-\omega)^v } = \sum_{r=v-1}^{\infty}
\binom{r}{v-1} \omega^{r-v+1},
\end{equation}

\noindent
and obtain

\begin{equation}
[A(\omega)]^{v} = \left[ \left( \frac{c-2}{c-1} \right) \frac{1}{N} \right]^v
\sum_{r=v-1}^{\infty} \binom{r}{v-1} \omega^{r+1}.
\label{eq:A^v}
\end{equation}

\noindent
Inserting $[A(\omega)]^{v}$ from Eq. (\ref{eq:A^v})
into Eq. (\ref{eq:gamma_s_omega}), we obtain

\begin{eqnarray}
L_s(\omega) &=&
\sum_{t=s}^{\infty}
\omega^t 
\sum_{v=s-2}^{t-2} 
(-1)^{v-s}
\binom{t-2}{v} \left[ \left( \frac{c-2}{c-1} \right) \frac{1}{N}   \right]^{v} 
\times
\nonumber \\
& & \ \ \ \ \ \ \ 
\sum_{m=s-2}^{\min\{v,N-2\}} 
m!
\Big\{
\begin{array}{l}
v  \\
m       
\end{array}
\Big\}
\binom{N-2}{m}
\binom{m}{s-2}.
\label{eq:gammasom}
\end{eqnarray}

\noindent
Finally, extracting the coefficients of $\omega^t$ from Eq. (\ref{eq:gammasom}), we obtain

\begin{eqnarray}
P_t(S=s) &=& 
\sum_{v=s-2}^{t-2} 
(-1)^{v-s}
\binom{t-2}{v} 
\left[ \left( \frac{c-2}{ c-1 } \right) 
\frac{1}{N} \right]^{v}
\times
\nonumber \\
& & \ \ \ \ \ 
\sum_{m=s-2}^{\min\{v,N-2\}} 
m!
\Big\{
\begin{array}{l}
v  \\
m       
\end{array}
\Big\}
\binom{N-2}{m}
\binom{m}{s-2}.
\label{eq:P_t(S=s)!}
\end{eqnarray}

\noindent
Note that this solution satisfies
$P_t(S=s)=0$ for $t<s$,
which serves as a quick sanity check.

The case of $s=N$ is of special importance in the context of the
cover time, because
$P(T_{\rm C} \le t) = P_t(S=N)$.
Inserting $s=N$ in Eq. (\ref{eq:gammaTaylor}),
we obtain

\begin{equation}
L_N(\omega) = \frac{\omega^2}{1-\omega} \frac{(N-2)!}{ \left( 1 + \frac{1}{A(\omega)} \right)_{N-2}}.
\label{eq:gamNom}
\end{equation}

\noindent
Expanding $L_N(\omega)$ from Eq. (\ref{eq:gamNom})
in powers of $\omega-1$, we obtain
 
\begin{eqnarray}
L_N(\omega) 
&\simeq&
\frac{1}{1-\omega}
- 2 - \frac{c-1}{c-2} (N-2) H_{N-2}
- \left\{ 1 + \frac{c-1}{c-2} (N-2) H_{N-2} 
\right.
\nonumber \\
&+& 
\left.
\frac{1}{2} \left( \frac{c-1}{c-2} \right)^2 (N-2)^2 
\left[ \left( H_{N-2} \right)^2 + H_{N-2}^{(2)} \right] \right\}
(\omega-1) 
\nonumber \\
&+& \mathcal{O} \left[ (\omega-1)^2 \right].
\label{eq:gNom}
\end{eqnarray}

\section{The conditional distribution $P(T=t|s)$ and its moments}

The probability that the RW has pursued $t$ steps, given that it has visited 
$s \ge 2$ distinct nodes, is given by

\begin{equation}
P(T=t|s) = \frac{ P_t(S=s) }{ \sum_{t=2}^{\infty} P_t(S=s) }.
\end{equation}

\noindent
Using the generating function $L_s(\omega)$, 
defined in Eq. (\ref{eq:gsomega})
this probability can be written in the form

\begin{equation}
P(T=t|s) = \frac{ P_t(S=s) }{  L_s(1) }.
\label{eq:Tts}
\end{equation}

\noindent
Inserting $L_s(\omega)$ from Eq. (\ref{eq:gammasw})
into Eq. (\ref{eq:Tts})  
we obtain

\begin{equation}
P(T=t|s) = \left( \frac{c-2}{c-1} \right) \left( 1 - \frac{s}{N} \right) P_t(S=s).
\label{eq:PTts}
\end{equation}

\noindent
The tail distribution $P(T>t|s)$ is given by

\begin{equation}
P(T>t|s) = \sum_{t'=t+1}^{\infty} P(T=t|s).
\label{eq:PTts7}
\end{equation}

The mean of the conditional probability distribution 
$P(T=t|s)$,
for $2 \le s \le N-1$,
is given by

\begin{equation}
\mathbb{E}[T|s] = \sum_{t=2}^{\infty} t P(T=t|s).
\label{eq:ETs}
\end{equation}

\noindent
Inserting $P(T=t|s)$ from Eq. (\ref{eq:Tts}) into Eq. (\ref{eq:ETs}),
we obtain

\begin{equation}
\mathbb{E}[T|s] =
\frac{1}{L_s(1)}
\sum_{t=2}^{\infty} t P_t(S=s).
\end{equation}

\noindent
Expressing the sum on the right hand side in terms of $L_s(\omega)$,
we obtain

\begin{equation}
\mathbb{E}[T|s] =
\frac{1}{L_s(\omega)}
\frac{\partial}{\partial \omega}
L_s(\omega) \bigg\vert_{\omega=1},
\end{equation}

\noindent
or in a more compact form

\begin{equation}
\mathbb{E}[T|s] =
\frac{\partial}{\partial \omega}
\ln
L_s(\omega) \bigg\vert_{\omega=1}.
\label{eq:ETsB}
\end{equation}

\noindent
Inserting $L_s(\omega)$ from Eq. (\ref{eq:gammasw}) into Eq. (\ref{eq:ETsB}),
we obtain

\begin{equation}
\mathbb{E}[T|s]
= 1 +  \left( \frac{c-1}{c-2} \right) (N-2) (H_{N-2}-H_{N-s-1}).
\label{eq:ETs1}
\end{equation}

\noindent
In the limit of $2 \le s \ll N$, Eq. (\ref{eq:ETs1}) can be approximated by

\begin{equation}
\mathbb{E}[T|s]
\simeq 1 + \left( \frac{c-1}{c-2} \right) (s-1),
\end{equation}

\noindent
which is consistent with the results presented in Ref. 
\cite{Debacco2015}.
In the opposite limit of 
$1 \le N-s \ll N$, we obtain

\begin{equation}
\mathbb{E}[T|s]
\simeq
1 +  \left( \frac{c-1}{c-2} \right) N (\ln N +\gamma - H_{N-s-1}).
\label{eq:ETs7}
\end{equation}

\noindent
Note that for $s=N$ the expectation value $\mathbb{E}[T|N]$ 
diverges. This is due to the fact that $s=N$ is an absorbing state,
namely once the system reached the state in which the RW has
covered all the $N$ nodes it will remain in this state forever.

The variance of the conditional distribution $P(T=t|s)$ is given by

\begin{equation}
{\rm Var}\left[ T|s \right] =
\mathbb{E}[T^2|s] - \left( \mathbb{E}[T|s] \right)^2.
\end{equation}

\noindent
It can be expressed in terms of the generating function $L_s(\omega)$,
in the form

\begin{equation}
{\rm Var}\left[ T|s \right] =
\mathbb{E}[T|s] + \frac{ \partial^2 }{\partial \omega^2 } \ln L_s(\omega) \bigg|_{\omega=1}.
\label{eq:VTs}
\end{equation}

\noindent
Carrying out the differentiation of $L_s(\omega)$,
which is given by Eq. (\ref{eq:gammaTaylor}), 
using Eqs. (\ref{eq:2F1lim})-(\ref{eq:d3F}),
we obtain

\begin{equation}
{\rm Var}\left[ T|s \right] 
= - N \left( \frac{c-1}{c-2} \right) (H_{N-2}-H_{N-s-1})
+ \left( \frac{c-1}{c-2} \right)^2 N^2 [ H_{N-2}^{(2)} - H_{N-s-1}^{(2)} ].
\label{eq:VTsb}
\end{equation}

\noindent
In the limit of $2 \le s \ll N$ the
variance can be simplified to

\begin{equation}
{\rm Var}\left[ T|s \right] =
\frac{c-1}{(c-2)^2}
(s-1) + \frac{c(c-1)}{2(c-2)^2 N}
(s-1)(s+2) + \mathcal{O} \left( \frac{1}{N^2} \right).
\end{equation}

\noindent
In the opposite limit of $N-s \ll N$,
it takes the form

\begin{eqnarray}
{\rm Var}\left[ T|s \right] 
&=&
\left( \frac{c-1}{c-2} \right)^2 N^2 
\zeta(2,N-s)
- \left( \frac{c-1}{c-2} \right) N \ln N  
\nonumber \\
&+& \left( \frac{c-1}{c-2} \right) N
(H_{N-s-1}-\gamma)     
- \left( \frac{c-1}{c-2} \right)^2 \left[ N + \frac{3}{2(c-1)} \right]
\nonumber \\
 &+& \mathcal{O} \left( \frac{1}{N} \right),
\end{eqnarray}

\noindent
where  $\zeta(m,n)$ is the Hurwitz zeta function
\cite{Olver2010}.

Summarizing the results of this Appendix, it was found that in the limit of
$2 \le S \ll N$ there is a linear relation between $\mathbb{E}[T|s]$
and $s$ and the variance of $P(T=t|s)$ is small.
Thus, in this limit the number of distinct nodes visited by
an RW is a good predictor for the elapsed time.
This property breaks down for large values of $s$, where $1 \le N-s \ll N$.
In this limit the mean $\mathbb{E}[T|s]$ saturates and the variance
$\mathbb{V}[T|s]$ becomes very large.
As a result, knowing the number of distinct nodes visited by an RW
provides little information about the elapsed time $t$.

\section{The moments of $P_t(S=s)$}

The $r$th moment of the distribution 
$P_t(S=s)$ at time $t \ge 2$ is given by

\begin{equation}
\langle S^r \rangle_t = \sum_{s=2}^{t} s^r P_t(S=s).
\end{equation}

\noindent
Below we calculate the moments of $P_t(S=s)$,
using the generating function $G_t(x)$,
defined by Eq. (\ref{eq:Gtxdef}).
Taking the derivative of $G_t(x)$ with respect to $x$
and setting $x=1$, we obtain

\begin{equation}
\langle S \rangle_t  = \frac{\partial}{\partial x} G_t(x) \bigg\vert_{x=1} .
\end{equation}

\noindent
Taking the second derivative and setting $x=1$ yields 

\begin{equation}
\langle S (S-1) \rangle_t   = \frac{\partial^2}{\partial x^2} G_t(x) \bigg\vert_{x=1} ,
\end{equation}

\noindent
where $\langle S (S-1) \rangle_t$ is the second factorial moment of $P_t(S=s)$.
In general, taking the $n$th derivative and setting $x=1$,
we obtain

\begin{equation}
\langle S(S-1) \dots (S-n+1) \rangle_t =
\frac{\partial^n}{\partial x^n} G_t(x) \bigg\vert_{x=1},
\label{eq:dnGt1}
\end{equation}

\noindent
which is the $n$th factorial moment.
The ordinary moments can be expressed in terms of the factorial moments 
in the form

\begin{equation}
\langle S^r \rangle_t =
\sum_{n=0}^{r} 
\Big\{
\begin{array}{l}
r  \\
n       
\end{array}
\Big\}
\langle S(S-1) \dots (S-n+1) \rangle_t,
\end{equation}

\noindent
Therefore, the $r$th moment is given by

\begin{equation}
\langle S^r \rangle_t =
\sum_{n=0}^{r} 
\Big\{
\begin{array}{l}
r  \\
n       
\end{array}
\Big\}
\frac{\partial^n}{\partial x^n} G_t(x) \bigg\vert_{x=1}.
\label{eq:SrtG}
\end{equation}

\noindent
For the analysis below, it is useful to define the generating function of the factorial moments,
which is given by

\begin{equation}
K_n(\omega) = \sum_{t=2}^{\infty}
\omega^t \langle S(S-1) \dots (S-n+1) \rangle_t.
\end{equation}

\noindent
Using Eq. (\ref{eq:dnGt1}), this generating function can be expressed in the form

\begin{equation}
K_n(\omega) = \sum_{t=2}^{\infty}
\omega^t \frac{\partial^n}{\partial x^n} G_t(x) \bigg\vert_{x=1}.
\label{eq:FMnom}
\end{equation}

\noindent
We also define the moment generating function

\begin{equation}
M_r(\omega) = \sum_{t=2}^{\infty}
\omega^t \langle S^r \rangle_t.
\label{eq:Mrom1}
\end{equation}

\noindent
Inserting $\langle S^r \rangle_t$ from Eq. (\ref{eq:SrtG})
into Eq. (\ref{eq:Mrom1})
and inverting the order of summations, we obtain

\begin{equation}
M_r(\omega) = 
\sum_{n=0}^{r} 
\Big\{
\begin{array}{l}
r  \\
n       
\end{array}
\Big\}
\sum_{t=2}^{\infty}
\omega^t
\frac{\partial^n}{\partial x^n} G_t(x) \bigg\vert_{x=1}
\label{eq:Mrom2}
\end{equation}

\noindent
The second sum in Eq. (\ref{eq:Mrom2}) is equal to $K_n(\omega)$,
given by Eq. (\ref{eq:FMnom}).
Therefore,

\begin{equation}
M_r(\omega) = 
\sum_{n=0}^{r} 
\Big\{
\begin{array}{l}
r  \\
n       
\end{array}
\Big\}
K_n(\omega).
\label{eq:Mrom3}
\end{equation}

\noindent
The generating function $G_t(x)$ can be written as a series expansion
around $x=1$, which takes the form

\begin{equation}
G_t(x) = \sum_{n=2}^{N}
\frac{ (x-1)^n }{n!}
\frac{\partial^n}{\partial x^n} G_t(x) \bigg\vert_{x=1}.
\label{eq:Gtxe}
\end{equation}

\noindent
Inserting $G_t(x)$ from Eq. (\ref{eq:Gtxe}) into Eq. 
(\ref{eq:Gamma_xo}), exchanging the order of the summations
and using Eq. (\ref{eq:FMnom}),
we obtain

\begin{equation}
GL(x,\omega) = 
\sum_{n=2}^N \frac{(x-1)^n}{n!}  K_n(\omega).
\label{eq:Gxo}
\end{equation}

\noindent
Comparing this sum term by term to Eq. (\ref{eq:Gammasol}),
one can express the generating function $K_n(\omega)$ 
in terms of $a_n(\omega)$ and $A(\omega)$.
In order to perform such comparison, we need to express the
factor of $x^2$ in Eq. (\ref{eq:Gammasol}) in terms of powers of $x-1$,
namely  $x^2 = 1 + 2(x-1) + (x-1)^2$.
Inserting this equality into Eq. (\ref{eq:Gammasol}) 
and rearranging terms
we obtain

\begin{eqnarray}
GL(x,\omega) &=&
\frac{ \omega^2 }{ 1 - \omega }
\left\{ 1 + \left[ 2 + \frac{N-2}{1 + \left( \frac{c-1}{c-2} \right) N \frac{1-\omega}{\omega}   } \right] (x-1)
\right.
\nonumber \\
&+&
\left.
\sum_{n=2}^{\infty} [ a_n(\omega) + 2 a_{n-1}(\omega) + a_{n-2}(\omega) ] (x-1)^n \right\}.
\label{eq:Gxom1}
\end{eqnarray}

\noindent
Comparing Eqs. (\ref{eq:Gxo}) and (\ref{eq:Gxom1}) term by term,
we obtain

\begin{equation}
K_0(\omega) = \frac{\omega^2}{1-\omega},
\end{equation}

\begin{equation}
K_1(\omega) = \frac{\omega^2}{1-\omega}
\left[ 2 + \frac{N-2}{1 + \left( \frac{c-1}{c-2} \right) N \frac{1-\omega}{\omega}   } \right]
\end{equation}

\noindent
and

\begin{equation}
K_n(\omega) = \frac{\omega^2}{1-\omega} n!
[ a_n(\omega) + 2 a_{n-1}(\omega) + a_{n-2}(\omega) ].
\end{equation}

\noindent
Going back from the factorial moments to the ordinary moments, we obtain

\begin{equation}
M_0(\omega) = \frac{\omega^2}{1-\omega},
\end{equation}

\begin{equation}
M_1(\omega) = \frac{\omega^2}{1-\omega}
\left[ 2 + \frac{N-2}{1 + \left( \frac{c-1}{c-2} \right) N \frac{1-\omega}{\omega}   } \right],
\label{eq:M1om}
\end{equation}

\noindent
and

\begin{eqnarray}
M_r(\omega) &=&
\frac{\omega^2}{1-\omega}
\left[ 2 + \frac{N-2}{1 + \left( \frac{c-1}{c-2} \right) N \frac{1-\omega}{\omega}   } \right]
\nonumber \\
&+&
\frac{\omega^2}{1-\omega}
\sum_{n=2}^r 
\Big\{
\begin{array}{l}
r  \\
n       
\end{array}
\Big\}
n! [ a_n(\omega) + 2 a_{n-1}(\omega) + a_{n-2}(\omega) ],
\label{eq:Mrom}
\end{eqnarray}

\noindent
for $r \ge 2$.
Expanding $M_1(\omega)$ from Eq. (\ref{eq:M1om}) in powers of $\omega$,
we obtain

\begin{equation}
M_1(\omega) = \sum_{t=2}^{\infty}
\left\{ 2 + (N-2) \left( 1 - \left[ 1 - \left( \frac{c-2}{c-1} \right) \frac{1}{N} \right]^{t-2} \right) \right\} \omega^t.
\label{eq:M1omt}
\end{equation}

\noindent
Comparing the coefficients of $\omega^t$ in Eqs. (\ref{eq:M1omt}) and
(\ref{eq:Mrom1}) term by term, we obtain 

\begin{equation}
\langle S \rangle_t = 
2 + (N-2) \left\{ 1 - \left[ 1 - \left( \frac{c-2}{c-1} \right) \frac{1}{N} \right]^{t-2} \right\},
\end{equation}

\noindent
which coincides with previous results
\cite{Tishby2021b}, 
obtained using different methods.
Inserting $r=2$ in Eq. (\ref{eq:Mrom}), we obtain

\begin{equation}
M_2(\omega) = \left( \frac{\omega^2}{1-\omega} \right)
\frac{4+2A(\omega)+A(\omega)N[5+2A(\omega)N]}{[1+A(\omega)][1+2A(\omega)]}.
\label{eq:M2om}
\end{equation}

\noindent
Inserting $A(\omega)$ from Eq. (\ref{eq:A}) into Eq. (\ref{eq:M2om}),
rearranging terms and expanding in powers of $\omega$, we obtain

\begin{eqnarray}
M_2(\omega) &=& 
\sum_{t=2}^{\infty}
\omega^t
\left\{ N^2 - (N-2)(2N-1) \left[ 1 - \left( \frac{c-2}{c-1} \right) \frac{1}{N} \right]^{t-2} \right.
\nonumber \\
&+& \left. (N-3)(N-2) \left[ 1 - 2 \left( \frac{c-2}{c-1} \right) \frac{1}{N} \right]^{t-2} \right\} . 
\end{eqnarray}

\noindent
Therefore, the
second moment is given by

\begin{eqnarray}
\langle S^2 \rangle_t &=&
N^2 - (N-2)(2N-1) \left[ 1 - \left( \frac{c-2}{c-1} \right) \frac{1}{N} \right]^{t-2}
\nonumber \\
&+& (N-3)(N-2) \left[ 1 - 2  \left( \frac{c-2}{c-1} \right) \frac{1}{N} \right]^{t-2}.
\end{eqnarray}

\noindent
Using the results for the first and second moments, we obtain the 
variance, which is given by

\begin{eqnarray}
{\rm Var}_t(S) &=&
(N-2) \left[ 1 - \left( \frac{c-2}{c-1} \right) \frac{1}{N} \right]^{t-2}
\nonumber \\
&-& (N-2)^2 \left[ 1 - \left( \frac{c-2}{c-1} \right) \frac{1}{N} \right]^{2(t-2)}
\nonumber \\
&+& (N-3)(N-2) \left[ 1 - 2 \left( \frac{c-2}{c-1} \right) \frac{1}{N} \right]^{t-2}.
\end{eqnarray}

\section{The generating function of $P(T_{\rm C}=t)$}

Inserting $s=N$ in Eq. (\ref{eq:P_t(S=s)!}), we obtain

\begin{equation}
P_t(S=N) = 
(N-2)!  
\sum_{v=N-2}^{t-2} 
(-1)^{v-N}  
\Big\{
\begin{array}{c}
v  \\
N-2       
\end{array}
\Big\}
\binom{t-2}{v} 
\left[ \left( \frac{c-2}{ c-1 } \right)   \frac{1}{N} \right]^{v}.
\label{eq:P_t(S=N)!}
\end{equation}

\noindent
In Eq. (\ref{eq:PTCtPtsN}) we identify the relation
$P(T_{\rm C} \le t) = P_t(S=s)$.
Therefore, the
probability mass function of the cover time is given by

\begin{equation}
P(T_{\rm C}=t) = P_t(S=N) - P_{t-1}(S=N).
\label{eq:PTCdiff}
\end{equation}

\noindent
The generating function of $P(T_{\rm C}=t)$ is given by
Eq. (\ref{eq:JPCT}).
Inserting $P(T_{\rm C}=t)$ from Eq. (\ref{eq:PTCdiff}) into Eq. (\ref{eq:JPCT})
and rearranging terms,
we obtain

\begin{equation}
J(\omega) = (1-\omega) L_N(\omega).
\label{eq:J_omega}
\end{equation}

In the analysis below we are interested in the limit $\omega \rightarrow 1^{+}$.
In this limit $A(\omega) \rightarrow - \infty$ and $1/A(\omega) \rightarrow 0^{-}$.
Thus, to analyze the denominator of 
Eq. (\ref{eq:gamNom})
we apply a small $a$ expansion of 

\begin{equation}
\frac{1}{(1-a)_m} = \frac{1}{\prod_{k=1}^{m} (k-a)} =   (a)_m.
\label{eq:1am}
\end{equation}

\noindent
Using Eq. (\ref{eq:1am}), the
small $a$ expansion of $1/(1-a)_m$ 
can be obtained from known results for $(a)_n$.
It is given by

\begin{equation}
\frac{1}{(1-a)_m} =
\frac{1}{m!} \left\{ 1 + H_m a + \frac{1}{2} \left[   (H_m)^2 - H_m^{(2)} \right] a^2
+ \mathcal{O} \left( a^3 \right) \right\},
\end{equation}

\noindent
where $H_m^{(2)}$ is the $m$th Harmonic number of the second order
\cite{Olver2010}.
Applying this expansion to the right hand side of Eq. (\ref{eq:J_omega}),
we obtain

\begin{eqnarray}
J(\omega) &=& 1 + \left[ 2 + \frac{c-1}{c-2} (N-2) H_{N-2}   \right] (\omega - 1)  
+ \left\{ 1 + \frac{c-1}{c-2} (N-2) H_{N-2} 
\right.
\nonumber \\
&+&
\left.
\frac{1}{2} \left( \frac{c-1}{c-2} \right)^2 (N-2)^2 \left[ (H_{N-2})^2 + H_{N-2}^{(2)} \right]  \right\}
(\omega - 1)^2   
\nonumber \\
&+& \mathcal{O} \left[ (\omega-1)^3 \right].  
\label{eq:Jwexp}
\end{eqnarray}

\noindent
Taking derivatives of $J(\omega)$ and setting $\omega=1$ yields the moments
of the distribution of cover times.

\noappendix

%\newpage
\end{document}